# Ultrashort Carbon Nanotubes with Luminescent Color Centers are Bright NIR-II Nano-Emitters

Somen Nandi[1,2]†, Quentin Gresil[1,2]†, Benjamin P. Lambert[1,2], Finn L. Sebastian[3], Simon Settele[3], Ivo Calaresu[4], Juan Estaun-Panzano[5], Anna Lovisotto[5], Claire Mazzocco[5], Benjamin S. Flavel[6], Erwan Bezard[5], Laurent Groc[4], Jana Zaumseil[3], Laurent Cognet[1,2,*]

[1] *Laboratoire Photonique Numérique et Nanosciences, Université de Bordeaux, 33400 Talence, France*

[2] *LP2N, Institut d'Optique Graduate School, CNRS UMR 5298, 33400 Talence, France*

[3] *Institute for Physical Chemistry, Heidelberg University, D-69120 Heidelberg, Germany*

[4] *Interdisciplinary Institute for Neuroscience, CNRS, Univ. Bordeaux, 33076 Bordeaux, France*

[5] *Univ. Bordeaux, CNRS, IMN, UMR 5293, F-33000 Bordeaux, France*

[6] *Institute of Nanotechnology, Karlsruhe Institute of Technology, Kaiserstraße 12, D-76131 Karlsruhe, Germany*

*Correspondence: laurent.cognet@u-bordeaux.fr

†These authors contributed equally to this work.

## ABSTRACT

In the fields of bioimaging, photonics, and quantum science, it is equally crucial to combine high brightness with nanoscale size in short-wave infrared (SWIR) emitters. However, such nano-emitters are currently lacking. Here, we report that when functionalized with luminescent color centers, ultrashort carbon nanotubes with length much shorter than 100 nm, are surprisingly bright in the near-infrared second-biological window (NIR-II) of the SWIR domain. We discuss the origin of this exceptional brightness based on the uncontrollable presence of quenching defects in dispersed carbon nanotubes. We further investigate the nonlinear photoluminescence behavior of color centers functionalized carbon nanotubes in response to varying excitation conditions, spanning from ensemble measurements to single-nanotube experiments. We discuss how this behavior influences the determination of their photoluminescence quantum yields, which can reach values as high as 20% for ultrashort ones detected at the single nanotube level. Notably, the corresponding NIR-II brightness exceeds that of well-known visible emitters, including quantum dots. After rendering them biocompatible, we demonstrate point-spread function engineering and high-resolution, 3-dimensional single-particle tracking using these bright ultrashort carbon nanotubes allowing nanoscale imaging in the NIR-II window within thick brain tissue.

## INTRODUCTION

Bioimaging has seen impressive and constant progress over the last decades regarding sensitivity and resolution at which biological structures can be studied *in vivo* [1–5]. Fluorescence microscopy is the most common technique, offering both single-molecule (-molecular) sensitivity and nanoscale (-super) resolution. To achieve such properties, creating small, bright and photostable fluorescent emitters was a major step forward [6–8], as were the remarkable advances in optical instrumentation, detectors and image analysis tools [1,2,4]. However, nanoscale bioimaging of living specimens remains limited to studying simple samples (isolated systems, monolayer cells, or small cell assemblies) for which bright visible emitters can be applied. To access thick, complex living tissues or *in vivo* specimens, high-resolution fluorescence microscopy needs to be transferred to the near-infrared (NIR) region, particularly to the NIR-II window (~1000-1350 nm), where tissues exhibit the best light penetration depth due to a combination of low light scattering and absorption [9,10]. Highly transmissive optical elements and low-noise sensitive detectors are now available in this wavelength range. However, despite intensive research, emitters with performances similar to those at visible wavelengths are still out of reach.

Since the discovery of their NIR photoluminescence (PL) covering the NIR-II window [11], single-walled carbon nanotubes (SWCNTs) have been envisaged as a promising platform for high-resolution *in vivo* imaging. Other approaches, such as NIR quantum dots or upconversion nanoparticles are also explored, but do not yet reach the aforementioned performance for single-molecule localization microscopy [10,12,13]. Accordingly, several proof-of-concept experiments based on SWCNTs have been proposed for bioimaging purposes, primarily for low-resolution small animal imaging [10,13] but also at the single nanotube level using SWCNTs for nanoscale imaging of local tissue (re-)organization [14,15] or for molecular sensing applications in live cells or tissue [16–18]. These achievements benefitted from SWCNTs' high brightness and photostability in the NIR-II window. Yet, luminescent SWCNTs must be longer than 100 nm for bright emission, which corresponds to the typical exciton diffusion range in such structures [19]. On the one hand, their nanoscale diameters enable access to some complex restricted structures, and their strong anisotropy turned out to be an asset in applications requiring unconventional mapping of some complex architectures [14,20,21]. On the other hand, lengths exceeding 100 nm are suboptimal for SWCNTs to serve as universal imaging or sensing probes that should simultaneously achieve all desired properties, namely brightness, photostability and tiny dimensions in all directions.

To significantly broaden their sensing applications, SWCNT should have their lengths reduced by one order of magnitude to reach a few tens of nanometers, comparable to the dimension of common imaging probes such as organic fluorophores, quantum dots (QDs) used in the visible range or circulating biomolecules, like IgGs[8]. This reduction should be achieved while maintaining their high brightness properties. Recently, the introduction of luminescent color centers on the sidewall of SWCNTs has led to the creation of nanotubes with enhanced luminescence properties [22–24]. These are referred here to as Color-Center bearing carbon NanoTubes (CCNTs). This development has also paved the way for bright ultrashort CCNTs (uCCNTs) [25,26], where mobile excitons can be trapped at local color center sites. When immobilized on surfaces, uCCNTs could be detected at the single molecule level but since then, there have been no reports on their application in bioimaging or sensing applications. The reasons for this are threefold. First, an efficient production of these nano-emitters in reproducible quantities is necessary for bioimaging applications. A second requirement is the preparation of stealth biocompatible uCCNTs with optimized brightness while retaining high colloidal stability. For such applications, phospholipid-polyethylene glycol (PLPEG) was extremely efficient [27] and has since been used routinely to prepare biocompatible SWCNT hybrids [14,15]. Finally, the most limiting factor was the anticipated loss of brightness of uCCNTs compared to their longer counterpart [28]. We demonstrate in the following that this is not the case. Instead uCCNTs are unexpectedly bright and can serve as unparalleled NIR-II nanoprobes in biological tissues.

## RESULTS AND DISCUSSION

**Synthetic Approach for Luminescent Color-centered Based uCCNTs and Their Characterization.** Creating uCCNTs from pristine SWCNTs requires two steps: nanotube shortening and covalent functionalization by luminescent color centers, which can be performed in any sequential order. Accordingly, we generated two sets of uCCNTs where the two steps are performed in reverse orders and using different types of luminescent color centers. The first synthetic strategy was adapted from[26] where $sp^3$ functionalization of the nanotube backbone is performed to introduce color centers, followed by chemical cutting at the functionalization sites. In short, uCCNTs were prepared by covalent functionalization of unsorted CoMoCAT (6,5)-SWCNTs with *p*-nitroaryl groups using aryldiazonium chemistry in oleum, followed by oxidative (chemical) cutting with hydrogen peroxide ($H_2O_2$) (Figure 1A, additional details are given in Method and Figure S1-5). A high yield is obtained during the production of uCCNT from the CCNT dispersion, as shown by only 50% decrease in $E_{11}$ absorbance peak (Figure S1A). These nanotubes will be denoted as ar-uCCNTs. The second preparation method involves shortening of chirality sorted (6,5)-SWCNTs by tip-sonication (mechanical cutting) followed by functionalization with luminescent oxygen defects involving a Fenton-like reaction using copper(II) sulfate ($CuSO_4(H_2O)_5$) and sodium-L-ascorbate, according to a recently reported protocol [29]. These nanotubes will be designated as ox-uCCNTs (Figure 1A).

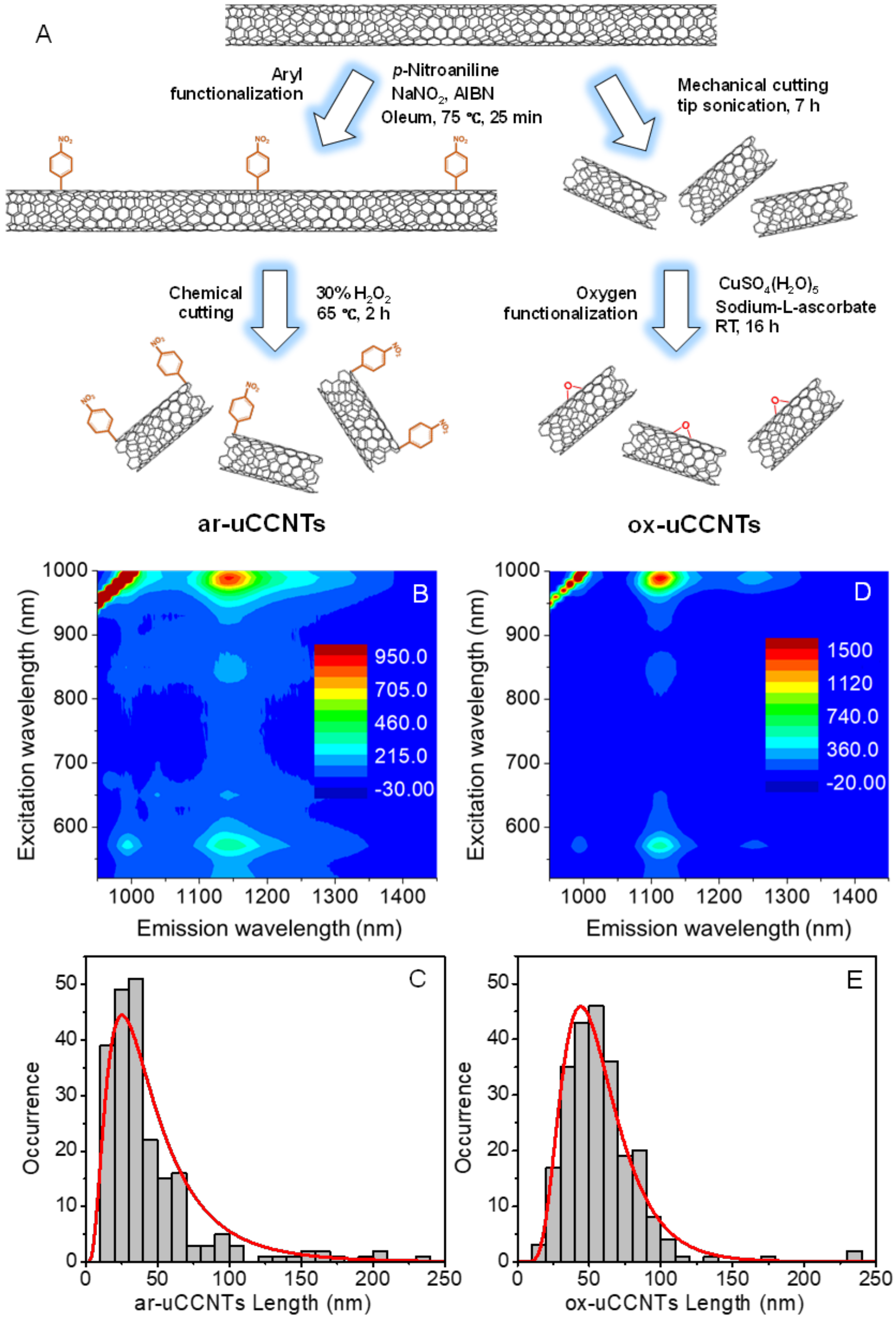

***Figure 1**. **Creation of ar- and ox-uCCNTs from pristine (6,5)-SWCNTs following two different approaches, where two steps - defect functionalization and shortening, are performed in reverse order, and using different functional groups as color centers.** Schematic of synthetic preparation (A). Two-dimensional PL map (excitation-emission profile) of (B) ar-uCCNTs and (D) ox-uCCNTs. AFM length distributions of (C) ar-uCCNTs (Median length = 40 nm, Mean length = 53 ± 40 nm) and (E) ox-uCCNTs (Median length = 54 nm, Mean length = 58 ± 27 nm), where more than 225 uCCNTs were measured for each sample.*

Surfactants like sodium deoxycholate (DOC) or sodium cholate are well known for preparing high-quality SWCNTs dispersions as they provide the brightest and most stable nanotube suspensions [30]. However, their utilization is incompatible with *in vivo* application due to their adverse effects on cell membrane integrity [31]. Biocompatible coatings, such as PLPEG, must be used to mitigate potential cytotoxicity concerns for bio-applications [14,15,32]. As the chemical cutting or mechanical cutting were performed in DOC or sodium dodecyl sulfate (SDS), respectively, surfactant exchange of uCCNTs with PLPEG was performed in a subsequent step (details are given in Method). Note that all the following characterization steps and experiments were carried out with these biocompatible dispersions, including *ex vivo* deep tissue imaging.

Spectroscopic measurements displayed in Figure 1, show $E_{11}$ emission peaks at ~985 nm at the band edge exciton of (6,5)-SWCNTs, as well as $E_{11}^*$ emission peaks at ~1140 nm for ar-uCCNTs (Figure 1B) and at ~1108 nm for ox-uCCNTs (Figure 1D). This difference in $E_{11}^*$ PL peak position arises from the difference in exciton trap depth for the two types of color centers (171 meV for aryl and 139 meV for oxygen). Atomic force microscopy (AFM) measurements performed on the two samples indicate median uCCNT lengths of 40 nm for ar-uCCNTs (Figure 1C, and Figure S6 and S8) and 54 nm for ox-uCCNTs (Figure 1E, and Figure S7-8). Next, we analyzed the PL quantum yields (PLQYs) of the two uCCNT dispersions and compared them to their longer counterparts denoted as ar-CCNTs for the long aryl-functionalized SWCNTs (i.e., before the last oxidation process) and ox-CCNTs for long oxygen-functionalized SWCNTs (i.e., not shortened by tip sonication before functionalization). The dispersion PLQY measurements for these four samples (ar-uCCNTs, ar-CCNTs, ox-uCCNTs and ox-CCNTs) were first performed metrologically using an integrating sphere (details are given in Method). For all four samples, the excitation was set near the $E_{22}$ optical transition with identical peak intensity (~30 W/cm²) using a pulsed supercontinuum laser. Mentioning this information is important because $E_{11}$* PL quickly becomes sublinear with excitation intensity, as demonstrated in previous studies.[33,34] Consequently, this nonlinear behavior significantly influences the PLQY values, which will be discussed in more detail below. The ar-uCCNTs and ar-CCNTs dispersions displayed very similar ensemble PLQY values (0.32% and 0.26%, respectively), as expected, considering that emission originates from local luminescent sites that were implanted on the long nanotubes before shortening (Table S1). Note that these values are on the same order as those of pristine long (6,5)-SWCNTs (~0.5-0.6%), whereas shortened pristine SWCNTs display significantly lower PLQY[29], indicating that enhanced PLQY values of uCCNTs are unambiguously attributable to covalent functionalization. Oxygen functionalized nanotubes, ox-(u)CCNTs, displayed superior average PLQYs compared to ar-(u)CCNTs, as previously observed [29], with values of 2.12% for ox-CCNTs and 0.70% for ox-uCCNTs (Table S1). Interestingly, these results indicate that the chemical cutting (in the case of ar-uCCNTs) does not create a very high number of additional structural defects (which act as quenching sites) as tip sonication does (in the case of ox-uCCNTs).

**Dependence of (u)CCNTs PLQYs (η) on Excitation Intensity and Excitonic Transition.** As mentioned above, it was previously observed that $E_{11}$* PL signals of CCNTs display sublinear behaviors on a wide range of excitation conditions both in ensemble measurement[33] (excitation intensity range of 1-50 kW/cm$^2$ near the $E_{11}$ phonon sideband) or at the single nanotube level[34] (excitation intensity range of 0.05-2 kW/cm$^2$ near $E_{11}$ or $E_{22}$ transitions). This behavior has been suggested to involve state-filling effects.[33] In general, the PLQY, $\eta$, is determined at a given excitation intensity, $I_{exc}$, from the measured PL signals, $S$, which can be written as $S = c\,\sigma \times \eta\, I_{exc}$, with $c$ representing both the collection and detection efficiency of the setup at the emission wavelength, and $\sigma$ is the absorption cross-section of the emitter. This implies that when the signal $S$ displays a sublinear behavior with $I_{exc}$, PLQY values are not constant and depend on the excitation conditions. The possibility to excite CCNTs at different wavelengths (such as at the $E_{11}$, $E_{22}$, or near phonon sidebands[24,33,34]) further complicates the determination of PLQYs as well as the comparison of reported values since they were generally obtained without taking into account these dependencies.

We therefore aimed to elucidate the PL behavior of CCNTs under varying excitation conditions. To this end, we first recorded the $E_{11}$* PL signals using both ensemble and single-nanotube measurements on a microscopy setup equipped for NIR-II imaging (see Method). This was done under excitation with a continuous-wave (cw) laser (at the $E_{11}$ transition) over a wide range of excitations intensities: from 20 mW/cm$^2$ to 50 W/cm$^2$ for ensemble measurements, and from 1 W/cm$^2$ to 3 kW/cm$^2$ for single-nanotube measurements. By normalizing ensemble and single nanotube measurements, we can display the PL dependence on a single graph (Figure 2A and S9). This approach highlights two distinct regimes. At intensities below ~5 W/cm$^2$, the PL is linear with excitation as expected at very low excitation levels. At higher intensities, a sublinear regime develops reaching a dependence of ${I_{exc}}^{0.8}$ up to 2 kW/cm$^2$.

The graph displayed in Figure 2A can be used to generate the dependence of PLQY on $E_{11}$ excitation intensity by dividing the reported PL values by $I_{exc}$ (Figure 2B and S9). At low excitation intensities, PLQYs remain constant and maximal (denoted as $\eta_0$), then a gradual decrease is observed with a $I_{exc}^{-0.2}$ dependence. Based on the values obtained from ensemble measurements using an integrating sphere, one can evaluate $\eta_0$ values of 2.5%, 0.8%, 0.3% and 0.4% for the suspensions of ox-CCNTs, ox-uCCNTs, ar-CCNTs and ar-uCCNTs, respectively. For easier applicability of this graph, we also indicated on the x-axis the number of excitons absorbed per unit of length and time, based on published values of $\sigma_{(6,5)E_{11}}$, the absorption cross-section per unit length of (6,5) nanotubes at their first order excitonic transition $E_{11}$.[35] Interestingly, the fact that the nonlinear regime begins at very low exciton densities compared to exciton lifetime (~100 ps)[29] and excursion range (~100-200 nm)[19] – indicates that the state-filling effect alone cannot account for this nonlinearity.

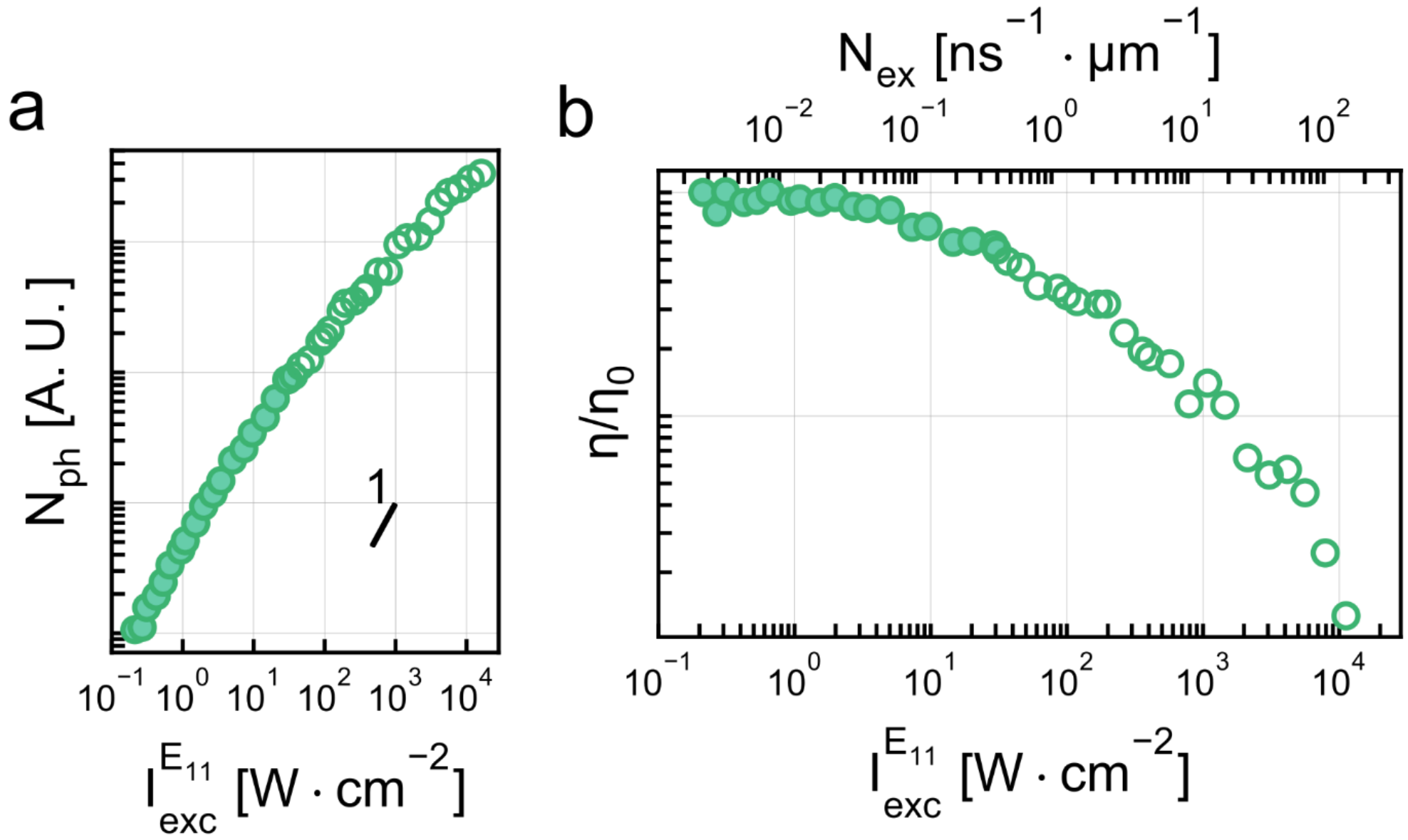

***Figure 2. PL Dependency on Excitation intensity.***
*(A) $E_{11}$* PL intensities as functions of cw excitation intensity ($E_{11}$ excitation), $I_{exc}^{E_{11}}$, obtained from ensemble (filled symbols) and single nanotube measurements (open symbols) with ox-CCNTs.*
*(B) The dependence of normalized PLQY ($\eta/\eta_0$) on $I_{exc}^{E_{11}}$ is obtained by dividing the measured PL values by $I_{exc}^{E_{11}}$, The corresponding exciton density $N_{exc}$ (in μm$^{-1}$·ns$^{-1}$) generated in nanotubes upon light absorption is shown on the top x-axis.*

Finally, we investigated whether the PLQY properties of CCNTs depend on the excitonic transition. Specifically, one might hypothesize that $E_{22}$ excitation promotes additional non-radiative exciton relaxation compared to $E_{11}$ excitation. To this end, we compared the absorption spectrum of the dispersions with their PL excitation (PLE) spectrum by monitoring the $E_{11}$* PL at low excitation intensity to ensure measurements remained within the linear regime. We observed that the PLE and absorption spectra display superimposable $E_{11}$ and $E_{22}$ peaks, indicating that PLQYs leading to emissive $E_{11}$* excitons are identical for both excitations (Figure S10).

Note that the definition of PLQY for $E_{22}$ excitation, depends on how the measurement is performed, as both $E_{11}$ mobile excitons (emitting below 1000 nm) and trapped $E_{11}$* excitons (emitting above 1100 nm) can in principle be detected. In cases where both types of excitons are considered, PLQY values will be larger by ~10% than if only trapped excitons are detected (see table S1).

**Single Molecule Microscopy Unveils Superior Brightness and PLQYs at The Single Nanotube Level.** To complement the above observations, we analyzed the $E_{11}$* PL brightness of the four samples at the single nanotube level under resonant $E_{11}$ excitation (Figure S11). We first recorded median PL signals as a function of excitation intensity for all samples and observed identical nonlinear behaviors (Figure 3A). We next present on Figure 3B the histogram of PL intensities at 200 $W/cm^2$. This confirms the overall superior brightness of oxygen-functionalized nanotubes compared to the aryl-functionalized nanotubes in agreement to the above observations at the ensemble level. The more striking and unexpected finding is the exceeding brightness of both types of uCCNTs compared to their longer counterparts, considering their relative lengths (Figure 3C).

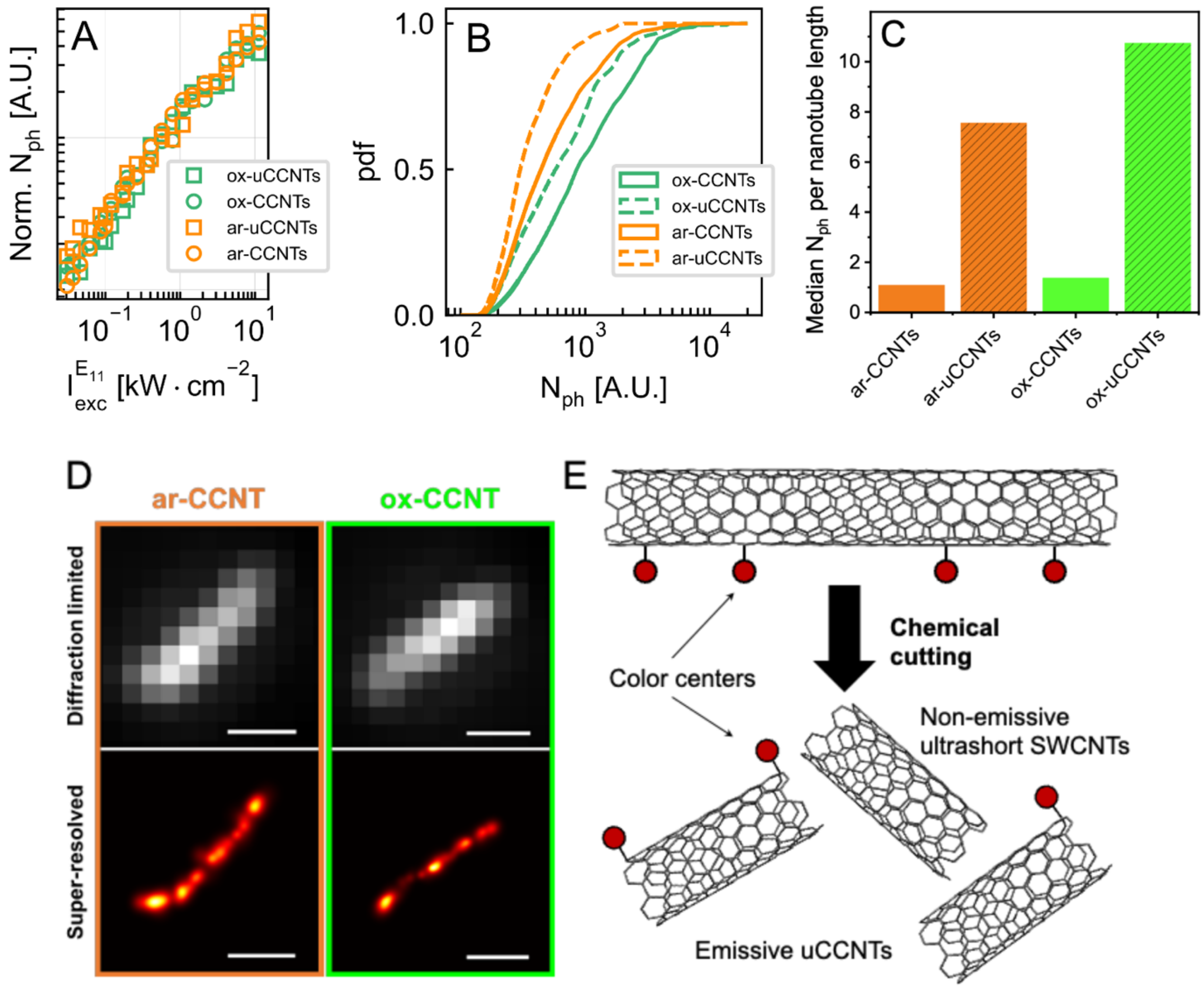

***Figure 3**. **Single-molecule brightness analysis and super-resolved images.***
*(A) Comparison of the measured PL intensity of all four samples (average signal of 8 single (u)CCNTs per condition) as a function of $E_{11}$ excitation intensity ($I_{exc}^{E_{11}}$) , indicating that the different samples follow identical non-linear behavior. (B) Cumulative distribution of the PL intensity of ar- and ox-uCCNTs with respect to their longer counterparts (ar- and ox-CCNTs), obtained from single nanotube images of $E_{11}$* PL emission ($I_{exc}^{E_{11}} = 200$ W/cm$^2$). More than 100 (u)CCNTs were analyzed per condition. (C) Comparison of measured median PL intensities normalized by nanotube median lengths ($I_{exc}^{E_{11}} = 200$ W/cm$^2$). (D) Diffraction limited (top) and super-resolution images (bottom) of the $E_{11}$* PL of long ar-CCNT (orange) and ox-CCNT (green) showing non-fluorescent segments. Scale bars: 1 µm. (E) Schematic illustration showing the partitioning of a CCNT into emissive uCCNTs and non-emissive ultrashort SWCNT segments after chemical cutting.*

Quantitatively, the brightness $B$ of a solution of emitters is defined at the ensemble level, by $B = \varepsilon \times \eta$, where $\varepsilon$ is the molar absorption coefficients of the emitter in solution, which has the unit of $M^{-1}.cm^{-1}$.[36] As mentioned previously, this metric is applicable in cases exhibiting sublinear PL behavior, where PLQY values consequently depend on the excitation intensity. At the single-molecule level, instead of the brightness $B$, it is more convenient to define the one-photon action cross-section as $A = \sigma \times \eta$ with $\sigma$ the absorption cross-section of an emitter which has a unit of $cm^2$. With these definitions, the one-photon action cross-section, $A$, also represents the brightness of the emitters but in a molecular unit. For convenience, the term 'brightness' can be used interchangeably to describe $A$ or $B$.

At the single nanotube level, we can thus write the fluorescence signals $S_i$ ($i$ refers to the 4 different samples) detected for each sample at the $E_{11}$* transition as $S_i = cA_iI_{exc}$, where $c$ represents the collection and detection efficiency of the setup at the emission wavelength, assumed to be identical for all samples and $A_i = \eta_i \times \sigma_{(6,5)E_{11}} L_i$, where $\eta_i$ is the PLQY of the given sample (eventually function of $I_{exc}$) and $L_i$ is the nanotube length. Since oxygen or aryl color centers do not modify the $\sigma_{(6,5)E_{11}}$ amplitude [22,37] and since $c \times I_{exc}$ is set identical for all measurements, one finds that $S_i \propto A_i \propto \eta_i L_i$. Figure 3B thus indicates that the PLQYs of individual uCCNTs are significantly larger than those of CCNTs. More precisely, one observes that while ar-uCCNTs (resp. ox-uCCNTs) and ar-CCNTs (resp. ox-CCNTs) differ in median length $L_i$ by one order of magnitude (by factors of 10.2 and 11.6 resp.; two-sample t-test, p-values = 0.05), their PL signals $S_i$ differs only by a factor of 1.5 (resp. 1.5) under identical excitation conditions. Considering the ratios of nanotube lengths, these measured PL brightness ratios indicate that the PLQYs of ar-uCCNTs and ox-uCCNTs are 6.9 times and 7.9 times higher, respectively, than those of ar-CCNTs and ox-CCNTs, as depicted in Figure 3C. These values are in conflict with the PLQY ratios measured at the ensemble level. We will demonstrate in the following that this discrepancy arises from the composition of uCCNT samples, which include both ultrabright functionalized uCCNTs and unfunctionalized ultrashort (and thus non-emissive) SWCNTs.

Structural or environmental quenching defects are well known to limit the PLQY of typical luminescent SWCNT samples [28,38,39]. The situation is different here in the case of ar- and ox-CCNTs, where PL emission benefits from exciton trapping at the luminescent centers to prevent the trapped $E_{11}$* excitons from encountering quenching defects [25,40], but also from increased oscillator strength of excitons in the zero-dimension quantum traps [37]. Sebastian *et al.* even indicated that a density of a few (4-8) defects per micrometer was preferable for optimum PLQY [41]. Accordingly, for the bright ox-CCNTs, the number of color centers per nanotube displayed a narrow unimodal distribution centered on 4-5 defects [42]. To unambiguously demonstrate that CCNTs contain non-emissive regions, we performed super-resolution imaging of luminescent sites [25,38] along ar- and ox-CCNTs. Figure 3D (Figure S12) exemplifies that for both ar- and ox-CCNTs, PL emission occurs from localized sites separated by dark intervals of varying extents (tens to hundreds of nm). Chemical cutting of CCNTs into segments having lengths shorter than these intervals should thus lead to two populations: non-emissive ultrashort SWCNTs and emissive uCCNTs displaying higher PLQYs than the original long CCNTs (Figure 3E). To further support this assumption, we estimated the relative concentrations of emitting uCCNTs and CCNTs by directly counting fluorescent entities under the single-molecule microscope (details are given in Method). We found that the concentration of uCCNTs is less than that of CCNTs divided by the length ratios of the two nanotube samples (Figure 4A). More quantitatively, ~88% (resp. 76%) non-luminescent SWCNT segments are present in the ar-uCCNT (resp. ox-uCCNT) dispersions.

The narrow unimodal distribution of the number of color centers implanted in ox-CCNTs [42] indicates that nearly all SWCNTs have been functionalized, and that the ensemble PLQY value of $\eta_0$ ~ 2.5% mentioned above adequately represents the PLQY of typical single ox-CCNTs.

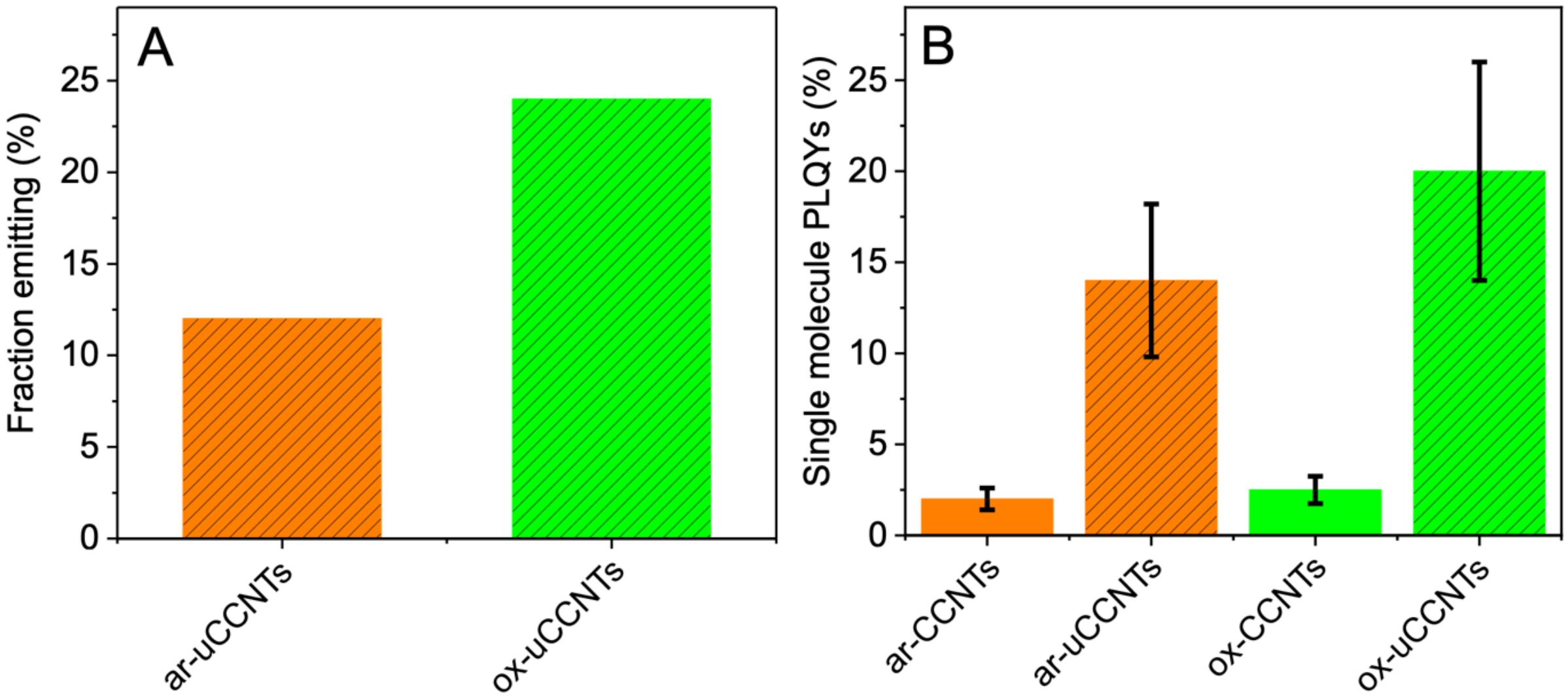

***Figure 4**. **Estimation of emitting ar- and ox-uCCNTs and PLQYs ($\eta_0$) calculated from the single nanotube analysis.** (A) Proportions of emitting ar- and ox-uCCNTs measured by single nanotube counting with respect to total ar- and ox-CCNTs. (B) Single nanotube PLQYs ($\eta_0$) of the four different samples.*

With this knowledge and from the measured PLQY ratios found at the single nanotube level (Figure 3C), we can deduce the PLQYs of the other three samples at the single-particle level. We found $\eta_0$ ~ 2% for ar-CCNTs, a value larger than the one at the ensemble level. This discrepancy suggests the presence of non-emissive nanotubes with ar-CCNTs in dispersion, likely resulting from the introduction of quenching defects during aryldiazonium chemistry in oleum. For uCCNTs, we found single nanotube PLQYs of $\eta_0$ ~ 14% for ar-uCCNTs and ~ 20% for ox-uCCNTs (Figure 4B). We estimate that the precision of these values is on the order of 30%, due to the inherent uncertainties in the ensemble PLQY measurement and the various correction factors applied, taking into account the experimental dispersion of data shown in Figure 2B. These values obtained for uCCNTs at the single-molecule level are significantly higher than previously reported values for nanotubes in metrological measurements [43–45].

Our results demonstrate that, due to the nonlinear PL properties of (u)CCNTs, careful consideration must be given to both the excitation intensities used for specific applications and the degree of nanotube functionalization. Both parameters will significantly impact the obtained PLQYs. For instance, in practical applications, single nanotube measurements are performed in the range of excitation intensities of 0.1-3 kW/cm$^2$, indicating that the effective PLQY relevant to single nanotube applications are half to one third of $\eta_0$ values.

Interestingly, the PLQY for ox-uCCNTs reaches the value of 18% as predicted for single oxygen quantum defects, expected to be observed in ideally clean oxygen-functionalized CCNTs (i.e., those free from quenching defects, including at its ends) [22,37]. Obtaining intrinsic PL properties for SWCNTs in dispersion is generally thought to be too difficult to achieve due to the inevitable generation of non-radiative defects during SWCNT synthesis and solvent dispersion process. However, our results show that this goal is within reach if we consider uCCNTs which exhibit near-optimal intrinsic PL properties regardless of the nanotube source, solubilization or shortening strategy, or functionalization method. These findings open up a broad range of potential applications in optical materials, quantum optics and bioimaging, for instance.

Regarding the latter, to date, QDs emitting at visible or far-red wavelengths are the prevailing single nanoparticle probes used in bio-imaging applications due to their high brightness *B*, ranging from 3x10$^5$ M$^{-1}$.cm$^{-1}$ up to 10$^6$ M$^{-1}$.cm$^{-1}$ (equivalent to one-photon action cross-sections of 1.10 up to 3.8x10$^{-15}$ cm$^2$ in single-particle units) [36]. However, QDs are strongly blinking nanoprobes despite extensive efforts to suppress such PL intermittency. In addition, at NIR-II wavelengths, their brightness strongly decreases in aqueous environments, preventing current bio-applications at the single-particle level [46–49]. The uCCNTs, therefore, represent a new opportunity for more photostable small single particle probes in the NIR-II window (Figure S13). Indeed, based on the PLQY measured above and using metrological determinations of $\varepsilon_{(6,5)E_{11}}$ (and $\sigma_{(6,5)E_{11}}$) values [35], they show brightness $B$, ranging from ~1.5-3x10$^{6}$ M$^{-1}$.cm$^{-1}$ (i.e. ~0.5-1x10$^{-14}$ cm$^{2}$ in single particle units) for ar-uCCNTs up to ~1-3x10$^{6}$ M$^{-1}$.cm$^{-1}$ (i.e. ~1-2x10$^{-14}$ cm$^{2}$) for ox-uCCNTs, equal to or even surpassing those of QDs. Importantly, beyond the absorption and PLQY properties, the much shorter luminescence lifetime of (u)CCNTs (~100's ps) [29] compared to QDs (>10's ns) and many other nano-emitters, can be an additional advantage as their PL saturation intensity is thus at much higher values.

**Single Particle Imaging Through Live Mice Brain Slices.** To illustrate the above advantageous properties, we first compared images of CdSe/ZnS QDs (with peak emission at 655 nm, additional details in Method) and uCCNTs immobilized on a glass surface when imaged with short integration time (50 ms) through live *ex vivo* mouse brain tissue slices of different thicknesses (Figure 5). As expected, single QD imaging becomes extremely challenging when increasing the imaging depth over 50 µm due to tissue autofluorescence and scattering at their red visible wavelengths and even impossible over 100 µm (Figure 5A). Accordingly, the localization precision of single QDs deteriorates rapidly as tissue thickness increases (from 8 to 44 nm precision for tissue thickness ranging from 0 to 50 µm, Figure 5B). In contrast, uCCNT imaging retains a high signal-to-noise ratio at such tissue thickness allowing excellent localization precision in all tissue thicknesses (i.e. precision ranging from 8 to 28 nm for tissue thickness ranging from 0 to 100 µm), due to the weak impact of tissue autofluorescence in the NIR II (Figure S14).

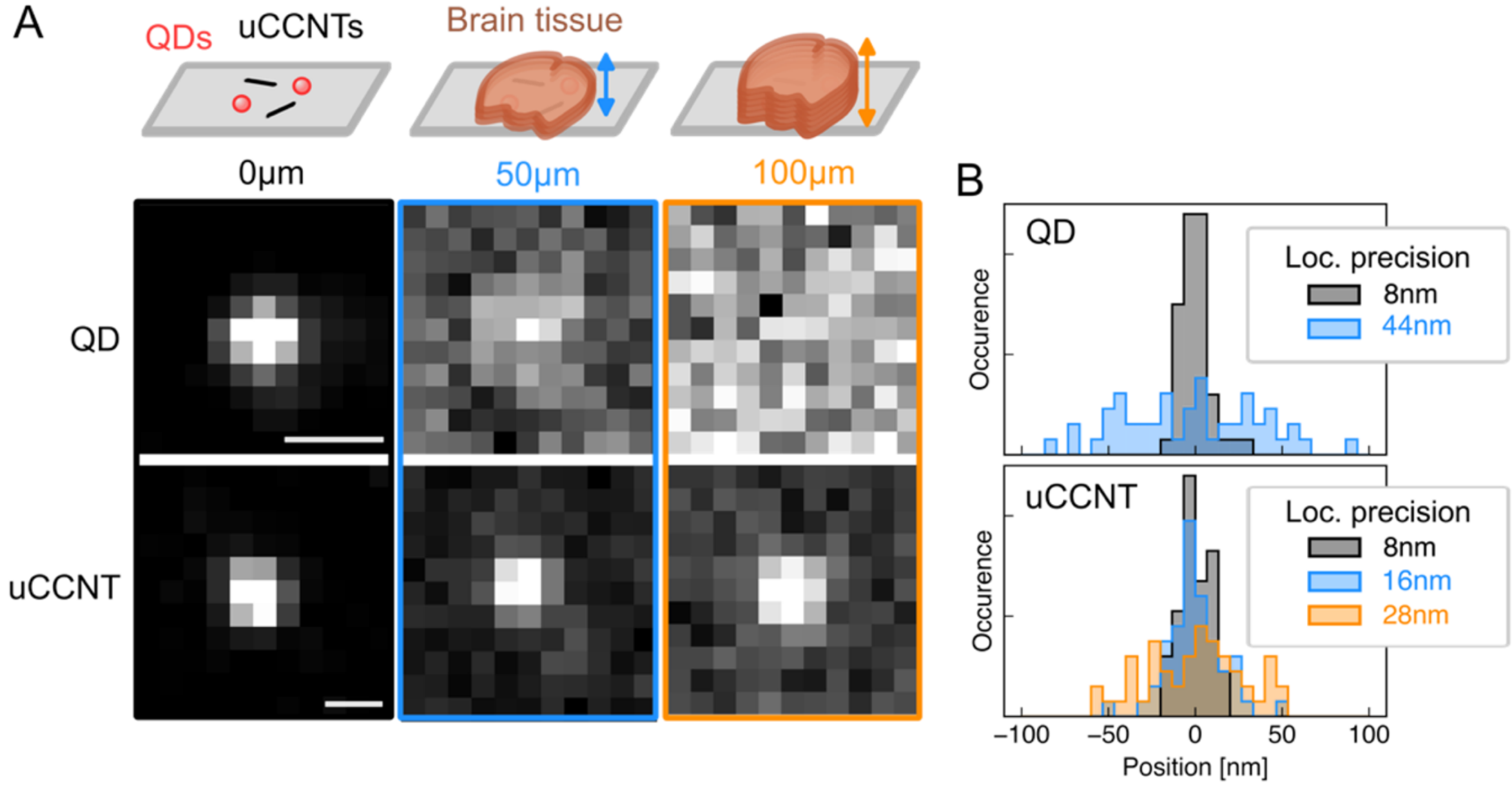

***Figure 5. Single-particle ex vivo imaging through mouse brain slices.*** *(A) Single-particle images of CdSe/ZnS QDs (655 nm emission) and ox-uCCNTs immobilized on a glass surface when imaged through live ex vivo mouse brain tissue slices of different thicknesses (0, 50 and 100 µm). Single QDs are no longer detectable at 100 µm due to tissue scattering and autofluorescence. Exposure time: 50 ms; Excitation power: 1 kW.cm$^{-2}$ at 488 nm and 2.4 kW.cm$^{-2}$ at 985 nm; Scale bars: 1 µm. (B) Corresponding single-particle localization precisions.*

We have finally applied uCCNTs to the 3-dimensional (3D) tracking of single particles deep in living biological tissue, one of the most demanding imaging approaches regarding nanoprobe brightness in high-resolution microscopy. Indeed, while single-particle tracking (SPT) is emerging in living biological tissues [50], the possibility of performing it in 3D has not yet been demonstrated. Here, we used organotypic rat brain slices and focused on the extracellular space (ECS) [51], details are given in Method.

Several approaches have been proposed to encode the 3D localization of single emitters with nanometer resolution, such as multiplane imaging, interferometric strategies or PSF engineering [52]. Importantly, they all require a high photon budget due to the necessity to redistribute the photon distribution, forming the image of a single emitter over wider spatial regions than the narrow Airy disc distribution obtained in ordinary microscopes. The availability of very bright NIR-II emitters is therefore essential, bearing in mind that detectors have poorer noise performance in the NIR-II than in the visible range. For this demonstration, we have chosen the Double-Helix PSF (DH-PSF) engineering [53] strategy because of its ability to achieve 3D super-localization with extended-depth-of-focus (i.e. over several micrometers), whereas most other methods are limited to axial dimensions of less than typically one micrometer.

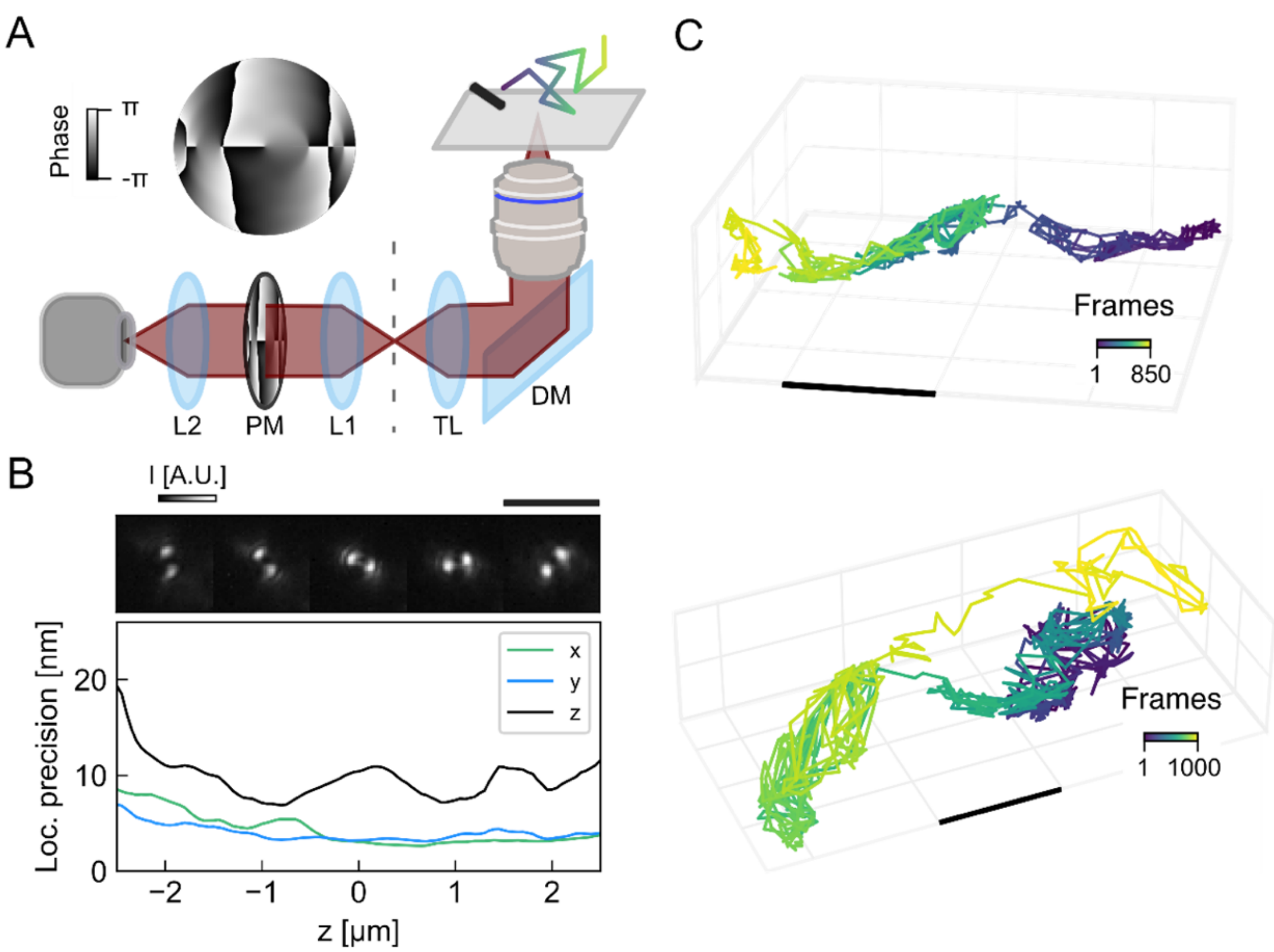

***Figure 6. Three-dimensional (3D) tracking of uCCNTs in brain tissue at depth.*** *(A) Schematic of setup for performing PSF engineering with a Double-Helix phase mask (PM). TL: tube lens; DM: dichroic mirror; L1 and L2: relay lenses. (B) Localization precision achieved with an immobilized ox-uCCNT. Exposure time: 30 ms; Scale bar: 16 μm. (C) Representative 3D trajectories of uCCNTs in organotypic rat brain slices at depth, resolving specific structures that were indiscernible in 2D. Exposure time: 30 ms; illumination intensity: 2.4 kW/cm²; Scale bar: 2 μm.*

Figure 6A, B shows the calibration of the 3D-localization performance of the NIR-II DH-PSF system along the extended-depth-of-focus using uCCNTs. When imaging immobilized ox-uCCNTs with only 30 ms integration time, we achieve single emitter super-localization with 5 nm lateral precision and 10 nm axial precision, which is better than $\lambda/100$ (and $\lambda/50$ respectively) at emission wavelength $\lambda = \sim 1110 - 1140\ nm$ (Figure 6B). The brightness of the nanoprobes as well as the reduced tissue scattering and autofluorescence at NIR-II wavelengths enables the transfer of these high-precision capabilities to tissue at depth. More quantitatively, localization precision of below 30 nm in all directions is obtained along 3D trajectories recorded in brain tissue slices at video rate (Figure 6B, C and Supplementary Movie S1). Most probably, the reduction in optical aberrations induced by brain tissue as the excitation wavelength increases towards the NIR region, as observed in multiphoton excitation fluorescence microscopy, must also contribute to maintaining localization precision of DH-PSF for 3D SPT in biological tissue. The combination of brightness, photostability, and large depth-of-focus allows for recording high content trajectories over tens of seconds at a high imaging rate. For instance, the analysis of trajectories shown in Figure 6C reveals patterns that would be invisible in 2D maps, such as hollow cylindrical shapes with a typical diameter of 200 nm.

## CONCLUSIONS

In conclusion, we have shown that uCCNTs display surprisingly bright near-infrared fluorescence compared to their longer counterparts. An explanation of this performance, supported by experimental evidence, is presented. We also investigated how the photoluminescence quantum yields depend on excitation conditions, spanning from ensemble measurements to single nanotube experiments. It is finally demonstrated that bright uCCNTs are promising nanoprobes for bioimaging in intact tissues, overcoming limitations associated with previously available emitters.

We posit that these novel NIR-II nanoprobes will enable in-depth investigations of diverse biological systems at the nanoscale, including *in vivo* specimens. Additionally, we also anticipate that bright infrared-emitting uCCNTs will find applications in various other domains, including optical materials or quantum technologies.

## EXPERIMENTAL METHODS

**Covalent Aryl Functionalization of SWCNTs and Chemical Cutting.** First, *p*-nitroaryl groups ($-C_6H_5NO_2$) were bound to the SWCNT sidewalls using diazonium chemistry in oleum following a previously described protocol [26]. A typical experiment consisted of dispersing 9 mg of raw unsorted CoMoCAT (6,5)-SWCNTs (≥95% as carbon nanotubes, CoMoCAT SG65i, Sigma-Aldrich) in 30 mL of oleum (20% free $SO_3$, Sigma-Aldrich) with magnetic stirring at room temperature for 3 h. 2.6 mg of sodium nitrite ($NaNO_2$; analytical standard, Sigma-Aldrich) were added to the SWCNTs/oleum dispersion followed by addition of 5.2 mg of *p*-nitroaniline (analytical standard, Sigma-Aldrich) (molar ratio of SWCNT carbon relative to *p*-nitroaniline was 20:1). The mixture was heated to 75 °C using a hot plate, and 7.5 mg of azobisisobutyronitrile (AIBN, 98%, Sigma-Aldrich) were added to initiate the reaction under constant magnetic stirring. After ~25 min of stirring at 75 °C, the reaction mixture was removed from the hot plate and cooled down to room temperature. The reaction mixture was added carefully (slowly, dropwise) to 600 mL of a 2.0 M aqueous solution of sodium hydroxide (NaOH, 97%, Sigma-Aldrich) with constant stirring in an ice bath until the solution reached pH 8. Then, the neutralized SWCNT suspension was vacuum-filtered through a 100 nm pore size PVDV hydrophilic membrane. The filter cake was washed with water and ethanol several times to remove unreacted chemicals/impurities and then dried overnight under vacuum. These nanotubes are denoted as ar-CCNTs.

In the next step, 2 mg of ar-CCNTs obtained on the filter paper were re-dispersed in 1% w/v sodium deoxycholate (DOC, 98%, Sigma-Aldrich) in 10 mL deuterium oxide ($D_2O$, Sigma-Aldrich) by homogenization in a shear mixer at 9000 rpm for 15 min, followed by probe sonication (Misonix-XL 2000) with a 3.2 mm diameter microtip at 4 W output power for 30 min under ice cooling. 10 mL of hydrogen peroxide ($H_2O_2$, 30 wt.% in $H_2O$, Sigma-Aldrich) was added to the dispersion and mixed properly. The mixture was then magnetically stirred at 65 °C for 2 h [26]. The color of the reaction mixture became lighter after the cutting process due to the carbon consumption (Figure S1A). We measured the absorption spectra of the samples before and after oxidation and evaluated the yield from changes in the $E_{11}$ absorption band. We obtained a yield of 50% for converting ar-CCNTs into ar-uCCNTs (Figure S1).

**Surfactant Exchange (SE) of ar-uCCNTs From 1% w/v DOC to Phospholipid-poly(ethylene glycol), PLPEG.** 500 µL of ar-uCCNTs in DOC (*Absorbance = 0.1 at $E_{11}$ band with 0.3 cm cuvette*) are mixed with 500 µL of 2.5% of 18:0 $PLPEG_{5000}$ (1,2-distearoyl-sn-glycero-3-phosphoethanolamine-N-[methoxy(polyethylene glycol)-5000, Layson Bio. Inc.) and subsequently bath sonicated. Then, 1500 µL methanol (ACS spectrophotometric grade, ≥99.9%, Sigma-Aldrich) was added dropwise, mixed and sonicated for 1 h in bath sonication followed by overnight incubation at room temperature. At this point, the DOC on SWCNTs should be completely replaced by PLPEG. After that, the mixture was filtered using 0.5 mL 3 kDa filter device at 10000 g (Mini Spin, Eppendorf) for 8 min followed by 5 times washing with $H_2O$.

For the surfactant-exchanged dispersion of *p*-nitroaryl-functionalized SWCNTs (ar-CCNTs), first the direct suspensions of ar-CCNTs from the filter cake (following the reaction in oleum and subsequent rinsing with deionized water) in 1% DOC were prepared. Typically, 1 mg of long ar-CCNTs was dispersed by probe sonication using a 3.2 mm diameter microtip probe sonication (Misonix-XL 2000) at 0.4 W/mL for 30 min under ice cooling. Nanotube bundles and impurities were precipitated by centrifugation (Beckman Coulter, 50.2 Ti rotor) at 13000 rpm (18000g) for 30 min at room temperature. The top ~80% of the supernatant were carefully collected and stored at room temperature until further use. A surfactant exchange was performed to 0.5% PLPEG solution in similar manner mentioned in previous section.

The exchanged samples were stored at 4 °C until further use. The surfactant-exchanged dispersion was characterized and compared by PL spectra. A peak shift of ~ 5 to 7 nm for both $E_{11}$ and $E_{11}$* peaks in PL measurements was observed (Figure S2). These peak shifts in the PL spectra are the evidence of surfactant exchange (27).

**Selective Dispersion of (6,5)-SWCNTs by Aqueous Two-phase Extraction (ATPE).** Nearly monochiral (6,5)-SWCNTs were obtained from CoMoCAT raw material (CHASM SG65i-L58) by aqueous two-phase extraction (ATPE) as described previously [54]. Briefly, the raw material was dispersed in DOC (BioXtra) and mixed in a two-phase system of dextran ($M_W$ = 70 kDa, TCI) and poly(ethylene glycol) (PEG, $M_W$ = 6 kDa, Alfa Aesar). (6,5)-SWCNTs were separated employing a diameter sorting protocol by addition of sodium dodecyl sulfate (SDS, Sigma-Aldrich). Semiconducting and metallic SWCNTs were separated by addition of sodium cholate (SC, Sigma-Aldrich) and sodium hypochlorite (NaClO, Sigma-Aldrich). The purified (6,5)-SWCNTs in DOC were transferred to 1% (w/v) SDS by concentrating them in a pressurized ultrafiltration stirred cell (Millipore) with a 300 kDa $M_W$ cutoff membrane followed by addition of 1% (w/v) SDS in $H_2O$.

**Shortening of (6,5)-SWCNTs and Transfer to PLPEG.** 5 mL of (6,5) SWCNTs in 1% SDS (w/v) obtained by ATPE were adjusted to an optical density of 0.33 $cm^{-1}$ at the $E_{11}$ transition with ultra-pure water and shortened by tip-sonication for 7 h (Sonic Vibra Cell VCX500, amplitude 25%, 8s on 2s off) under constant cooling in a 5 °C water bath. After centrifugation at 60,000*g* for 30 min (Beckman Coulter Avanti J-26S XP), the obtained dispersion and 5 mL of a dispersion of pristine (6,5)-SWCNTs were functionalized and processed according to a previously reported protocol [29]. The introduction of defects is manifested by the appearance of $E_{11}$* emission (Figure S3) and increase of the Raman D-mode (Figure S4). In the following, the functionalized shortened and long nanotubes will be referred to as ox-uCCNTs and ox-CCNTs, respectively. The transfer to PLPEG was performed using an adapted procedure by Welsher *et al.* [27]. Dispersions of functionalized SWCNTs were mixed with 18:0 $PLPEG_{5000}$ (Avanti Lipids, 2 mg/mL), and dialyzed in a 1 kDa dialysis bag (Spctra/Por®, Spectrum Laboratories Inc.) for 7 days against ultra-pure water. The obtained dispersions were bath-sonicated for 15 min and concentrated *via* spin-filtration (Amicon Ultra-4, 10 kDa) prior to further use (Figure S5).

**UV-Vis-NIR Absorption.** Absorption spectra (400-1100 nm) of aryl-functionalized samples (ar-CCNTs and ar-uCCNTs) were recorded with a spectrophotometer (EVOLUTION 220, Thermo Scientific). A quartz cuvette (path length 3 mm) was used for absorption measurements. Baseline-corrected UV-Vis-NIR absorption spectra for ox-CCNTs and ox-uCCNTs were acquired using a Cary 6000i absorption spectrometer (Varian, Inc.) and a quartz cuvette with 10 mm path length

**Photoluminescence (PL) Spectra and Two-dimensional (2D) Excitation-emission PL Map.** PL spectra and 2D PL maps for pristine and aryl functionalized samples, ar-(u)CCNTs, were recorded with a NanoLog spectrofluorometer (HORIBA Scientific). The samples were excited with a 450 W xenon source dispersed by a double-grating monochromator. The slit widths of the excitation and emission beams were 10 nm. The PL spectra were collected using a liquid-$N_2$-cooled linear InGaAs array detector (Symphony II). The integration times for the PL spectra and 2D map were 2 and 5 s, respectively. Before the PL experiments, $Na_2S_2O_4$ (technical grade, Sigma-Aldrich) solution in $D_2O$ was added to quench residual $H_2O_2$, followed by adding a few microliters of $NaHCO_3$ (≥99.7%, Sigma-Aldrich) solution until the solution becomes weakly basic (pH ~ 8). PL spectra of pristine and oxygen-functionalized aqueous (6,5)-SWCNT dispersions (ox-CCNTs and ox-uCCNTs) were recorded with a Fluorolog FL-3 spectrometer (Horiba Jobin-Yvon) equipped with a 450 W xenon light source and a double-grating monochromator (excitation and emission beam slit width 14 nm). Emitted light was passed through a long-pass filter (850 nm), Czerny-Turner spectrograph and guided onto a liquid nitrogen cooled InGaAs line camera (Symphony II). In the 2D maps, the excitation wavelength is shown on the y axis, the emission wavelength is depicted on the x axis and the PL intensity is indicated by the colormap.

**Raman Spectra.** Raman spectra of the pristine and ar-(u)CCNTs samples were recorded using a quartz cuvette with a 638 nm excitation (XploRA spectrometer, HORIBA Jobin Yvon). The integration time was 6 s, and 10 accumulations were averaged for each sample. Covalent bonding of an aryl functional group to SWCNTs sidewalls introduces a $sp^3$ defect in the $sp^2$ carbon lattice. This defect causes a considerable perturbation in local electronic properties when coupling strongly to the lattice vibrations, and gives rise to a symmetry-breaking, defect-induced Raman scattering (D phonon, ~1300 $cm^{-1}$). The intensity of this Raman band with respect to the in-plane stretching mode of the $sp^2$-bonded carbon lattice (G band, ~1585 $cm^{-1}$) is roughly proportional to the degree of aryl functionalization. The $I_D/I_G$ ratio can be used to evaluate the luminescent color center density in functionalized SWCNTs [41]. In Figure S1B, we observed a significant increase in the ratio after functionalization (from 0.05 to 0.14). The new color center emission at ~1140 nm was accompanied by an increase in the Raman $I_D/I_G$ ratio of the functionalized (6,5)-SWCNTs, which confirms the covalent attachment of aryl groups to the $sp^2$ carbon lattice. After prolonged oxidation by $H_2O_2$, the $I_D/I_G$ ratio (~0.2) remains low implying no significant increase of total numbers of color centers after oxidation. Raman spectra of drop-cast films for pristine and ox-

(u)CCNTs were collected with an inVia Reflex confocal Raman microscope (Renishaw) in backscattering configuration (50× long working distance objective, Olympus, N.A. 0.5) under near-resonant excitation (532 nm). Over 3000 individual spectra were collected and averaged for each sample.

**Atomic Force Microscopy (AFM).** Atomic force microscopy (AFM) images of ox-(u)CCNTs were recorded with a Bruker Dimension Icon AFM under ambient conditions. For measurements of individual ox-uCCNTs to perform a statistical length analysis, $Si/SiO_2$ substrates were cleaned by bath sonication in acetone (10 min) and isopropyl alcohol (10 min), incubated with aqueous Poly-L-lysine (PLL) hydrochloride (Sigma-Aldrich) solution (0.1 g/L) for 10 min, rinsed with deionized (DI) water, and blow-dried with nitrogen. The dispersions were diluted to an optical density of 0.2 $cm^{-1}$ at the $E_{11}$ transition. A drop of the diluted dispersion was placed on the $Si/SiO_2$ substrate and incubated for 10 min. After removal of the drop, the substrate was carefully washed with ultra-pure water and blow-dried with nitrogen. For ox-CCNTs, the procedure was carried out without prior treatment of the $Si/SiO_2$ substrates with PLL. Similarly, ar-uCCNTs were imaged using mica surface and AFM probes with a tip radius of ~ 7 nm (AC160TS-R3 AFM probe from Oxford Instruments) with in the AC-tapping mode (Oxford Instruments, MFP-3D Infinity, Asylum research) at a sampling rate of 1.0 Hz and a 512-line resolution. For ar-CCNTs, the procedure was carried out without prior treatment of the substrates with PLL.

We measured more than 225 nanotubes for each uCCNTs sample and more than 275 nanotubes for each CCNTs sample to analyze the length distributions using Gwyddion software (Figure S6-8). A two-sample t-test at the $p = 0.05$ significance level indicates that the difference in population mean values between the long nanotubes (CCNTs) and their ultrashort counterparts (uCCNTs) is statistically significant.

**Determination of Photoluminescence Quantum Yields (PLQYs) in Ensemble Measurement.** Photoluminescence quantum yields (PLQYs), $\eta$, of ar- and ox-(u)CCNTs were determined (Table S1) *via* an absolute method from the number of emitted ($N_{em}$) and absorbed ($N_{abs}$) photons as reported previously [55]:

$$\eta = \frac{N_{em}}{N_{abs}}$$

CCNT dispersions were diluted to an optical density of 0.15-0.2 $cm^{-1}$ at the $E_{11}$ transition to minimize reabsorption. A quartz cuvette (Hellma Analytics QX) with 1 mL of the analyte was placed in an integrating sphere (LabSphere, Spectralon coating). The output of a pulsed supercontinuum laser (NKT Photonics SuperK Extreme) was spectrally filtered to select the wavelength of the $E_{22}$ transition (570 nm for (6,5)-nanotubes) and the laser beam was guided into the integrating sphere. The excitation intensity was set to ~30 $W/cm^2$, close to the linear regime of emission of the CCNTs (see the main text). The light exiting the integrating sphere was coupled into a spectrometer (Acton SpectraPro SP2358) with an optical fiber, and spectra were acquired with a liquid nitrogen-cooled InGaAs line camera (Princeton Instruments OMA V:1024). The absorption of the sample was calculated from the difference in the attenuation of the laser light (at 570 nm) compared to the pure solvent. Similarly, the number of emitted photons was calculated from the signal difference in the spectral range from 900-1400 nm for the sample and solvent. The sensitivity of the detector and the absorption of optical components was accounted for by correction with a lamp of known spectral output (Thorlabs SLS201/M, 300-2600 nm).

**Single Molecule Microscopy: PL Imaging at Single Nanotube Level for Brightness Analysis.** CCNTs (surfactant exchanged to PLPEG) PL imaging was carried using a custom-built inverted microscope (Nikon Eclipse Ti) equipped with a 60X water immersion (1.27 NA, Nikon) objective.

A 988 nm laser (AeroDIODE) was used to resonantly excite the functionalized (6,5)-nanotubes at their first order excitonic transition ($E_{11}$). The incident laser beam was circularly-polarized to ensure that CCNTs are excited irrespective of their in-plane orientation. To produce wide-field images of individual nanotubes, PL from ar- and ox-(u)CCNTs was collected with the same objective and imaged on an InGaAs camera (C-RED 2, First Light Imaging). A dichroic mirror (Di02-R1064, Semrock, Rochester, NY, USA) was used to guide the excitation light onto the sample to illuminate the sample and the long-pass emission filters (RazorEdge 1064, Semrock, Rochester, NY, USA) to block reflected excitation light and to detect the PL from the $E_{11}^*$ state. Images of ar- and ox-(u)CCNTs were recorded at 10 ms exposure time. To enable imaging of individual immobilized (u)CCNTs while in solution, the following protocol was used: Standard microscope glass coverslips were initially treated with plasma cleaning (180 W power, for a duration of 2 minutes) to eliminate surface contaminants and enhance their surface hydrophilicity. The cleaned coverslips were then incubated with an aqueous solution of PLL (100 μL, 0.1 g/L) for 30 min to promote electrostatic adhesion of the nanotubes. A 50 μL aliquot of a diluted (u)CCNT dispersion was subsequently incubated onto the PLL-coated surface for 1 h to enable adsorption of the nanotubes. Following adsorption, the surfaces were washed with DI water to remove non-adherent material and were allowed to dry overnight at ambient temperature. Prior to imaging, the prepared coverslips were mounted in a Ludin chamber, and a drop of DI water was added to the sample to maintain hydration during image acquisition (Figure S15).

**Estimation of Proportions of Emitting ar- and ox-uCCNTs With Respect To Total ar- and ox-CCNTs Measured by Single Nanotube Counting.** We estimated the relative numbers of CCNTs and uCCNTs by directly counting fluorescent entities under the single molecule microscope. To do so, we first estimated the total concentration of nanotube dispersion from

absorption spectra. To determine the correct $E_{11}$ absorbance value for ar-(u)CCNTs samples, we first needed to remove the background caused by the scattering of carbonaceous materials [56]. We applied a fitting function in the 400 to 800 nm range, using the formula $b + \frac{k}{\lambda^c}$ where λ is the wavelength, and b is an offset manually set to -0.015 (Figure S16A). After subtracting the measured background from the overall absorbance spectrum, we fitted the absorption spectra using six Voigt peaks in the 800 to 1100 nm spectrum range (Figure S16B). The absorbance peaks for $E_{11}$ at ~985 nm were then evaluated for both ar-CCNTs and ar-uCCNTs, with a value of 0.19 in both cases. We did not need to subtract the absorbance background for ox-(u)CCNTs samples as the background was negligible for purified/sorted (6,5)-SWCNTs. The $E_{11}$ absorbance peaks were 0.4 and 0.24 for ox-CCNTs and ox-uCCNTs, respectively. With these values, we estimated the proportion of emitting uCCNTs with respect to total CCNTs in the following paragraph. First we calculated the total number of CCNTs using the formula:

$$N_{CCNTs} = \frac{C_{CCNTs}\, V N_A}{N_C\, L_{CCNTs}}$$

Where, $N_{CCNTs}$ is the total number of CCNTs, $C_{CCNTs}$ is the total concentration of carbon atoms in CCNTs that is directly proportional to $E_{11}$ absorption value, $V$ is the total volume (constant), $N_A$ is the Avogadro number (constant), $N_C$ is the number of carbon per micron length of CCNT (constant) and $L_{CCNTs}$ is the median length of CCNTs. The ratio of total number of CCNTs and uCCNTs can be easily written as follows:

$$\frac{N_{CCNTS}}{N_{uCCNTs}} = \frac{C_{CCNTs}\, L_{uCCNTs}}{C_{uCCNTs}\, L_{CCNTs}}$$

This leads to the value of ~0.10 for aryl functionalized sample and ~0.15 for oxygen functionalized samples. Considering $N_{uCCNTs} = N_{uCCNTs}^{emissive} + N_{uCCNTs}^{non-emissive}$, we assume all CCNTs are emissive (as indicated by Raman data on ox-CCNTs from [42]), thus we have: $N_{CCNTs} = N_{CCNTs}^{emissive}$. From the single-molecule microscopy experiments, we obtained $N_{CCNTs}^{emissive} = p\, N_{uCCNTs}^{emissive}$, where p is determined. As a result, we estimated that ~12% luminescent SWCNTs (~88% non-luminescent resp.) are present in the ar-uCCNT dispersions. In similar manner, we estimated the proportion of luminescent SWCNT for ox-uCCNTs and observed to be 24% (~76% non-luminescent resp.).

**Single Particle Imaging Through Live Mice Brain Slices.** A mixture of uCCNTs and biocompatible CdSe/ZnS QDs were immobilized on a coverslip by spin-coating. We used QDs having a peak emission at 655 nm that were passivated by a PEG layer, and decorated with F(ab')2 IgG (H+L, goat anti-rabbit) fragments (Qdot™ 655, Q11422 MP, Invitrogen). Mice (strain C57BL6/JRj purchased from Janvier laboratories and reared at the University of Bordeaux animal facilities) were euthanized by cervical dislocation and whole brains were swiftly extracted. Coronal sections (50 to 100-µm thick) were prepared in a VT1200S vibratome (Leica) in ice-cold Phosphate Buffered Saline (PBS) solution and left to recover for at least 20 min at room temperature. Experiments were performed following the European Union directive (2010/63/EU) on protecting animals used for scientific purposes. They were approved by the Ethical Committee of Bordeaux University (CEEA 50, France) and the Ministry of Education and Research under license number ##32540-2021072016125086. Slices of the desired thickness were then placed on the coverslip using an electrophysiology harp. Images were collected at 37 °C in a 3D-printed chamber with controlled temperature. Oxygenated warmed aCSF was perfused throughout data collection by a peristaltic pump. Imaging was conducted using a customized inverted Nikon microscope (Eclipse Ti) equipped with a 60X water immersion objective (Olympus Plan Apo XLPlan N) with a numerical aperture (N.A.) of 1.05. The uCCNT imaging was carried out using a 985 nm (2.4 $kW.cm^{-2}$, Oxxius) excitation laser and an InGaAs camera (C-RED 2, First Light Imaging). A dichroic mirror (Di02-R1064, Semrock, Rochester, NY, USA) and long-pass emission filter (RazorEdge 1064, Semrock, Rochester, NY, USA) were used to illuminate the sample and detect NIR PL from uCCNTs. On the other hand, QDs were excited at 488 nm (1 $kW.cm^{-2}$, Coherent) and detected using a sCMOS camera (PrimeBSI, Princeton Instrument). A visible dichroic mirror (FF506-Di03-25x36, Semrock, Rochester, NY, USA) and long-pass emission filter (FF01-655/15-25, Rochester, NY, USA) were employed. 50 frames were recorded for both QD and uCCNT and localization precision was measured as the standard deviation of the 50 measured positions after Gaussian fitting from ThunderSTORM[57] using default fitting parameters.

**Organotypic Rat Brain Tissue Preparation for Imaging.** Organotypic slice cultures were prepared as previously described [58]. Briefly, 350 μm hippocampal slices were obtained from postnatal day 5 (P5) Sprague-Dawley rats using a McIlwain tissue chopper, and were placed in cold (4 °C) dissection medium containing 175 mM sucrose, 25 mM D-glucose, 50 mM NaCl, 0.5 mM $CaCl_2•2H_2O$, 2.5 mM KCl, 0.66 mM $KH_2PO_4$, 2 mM $MgCl_2•6H_2O$, 0.28 mM $MgSO_4•7H_2O$, 0.85 mM $Na_2HPO_4•12H_2O$, 2.7 mM $NaHCO_3$ and 2 mM HEPES; ~320 mOsm, pH 7.3 (all products were from Sigma Aldrich, unless otherwise specified). After 25 min of incubation, slices were transferred onto white hydrophilic polytetrafluoroethylene membranes (0.45 μm; Millipore FHLC) set on Millicell Cell Culture Inserts (Millipore, 0.4 mm; 30 mm diameter), and cultured for up to 14 days on multiwell-plates at 35 °C/5% $CO_2$ in a culture medium composed of 50% Basal Medium Eagle, 25% Hank's balanced salt solution 1X (with $MgCl_2$/with $CaCl_2$), 25% heat-inactivated horse serum, 0.45% D-glucose and 1 mM L-glutamine (all products were from Gibco, unless otherwise specified). The medium was changed every 2-3 days.

**Organotypic Rat Brain Tissue Incubation.** The tissue slices were first incubated with uCCNTs (final concentration: 0.2 mg/L at $E_{11}$ absorption) for 1 h at 35 °C / 5% $CO_2$ in culture medium and then the slices were placed in a Ludin chamber for image acquisition up to 1 h after rinsing in warm recording solution for 10 min. The recording chamber was filled with HEPES-based artificial cerebrospinal fluid (aCSF) solution (145 mM NaCl, 4 mM KCl, 2 mM $CaCl_2.2H_2O$, 1.0 mM $MgCl_2.6H_2O$, 10 mM HEPES and 10 mM D-glucose; ~310 mOsm, pH 7.35) at 35 °C with controlled temperature (Tokai Hit). An electrophysiology harp was used to maintain the slices at the bottom of the chamber and evenly placed regarding the focal plane, which facilitated the image acquisition. Before recording, transmission white light and a low magnification objective (20X, Nikon Plan Fluor, 0. 5 N.A.) were used to evaluate their quality, to check the position in the entire slice and determine the brain region to be imaged (Figure S14). To avoid non-physiological data acquisition, the first 10 μm of tissue layer was always discarded to exclude the first cell layer, which was potentially damaged during slicing.

**Three-dimensional (3D) Tracking of uCCNTs in Rat Brain Organotypic Slices.** 3D single-particle tracking (SPT) was performed with the same microscope set-up described in section 1.3, with the addition of an imaging relay consisting of two achromat doublet lenses placed after the microscope. The double-helix phase mask (Double Helix Optics) was inserted at the 4f Fourier plane, conjugated to the back focal plane of the objective. The double-helix phase mask was placed on a kinematic mount with ~ 1 μm precision to ensure correct alignment. An InGaAs camera (C-RED 2, First Light Imaging) was employed for image capture at 30 ms exposure time. For Double-Helix PSF (DH-PSF) calibration, uCCNTs were immobilized on a coverslip by spin-coating. A calibration z-stack is acquired with 50 nm steps of 50 images covering the full axial range of the DH-PSF. Phase retrieval of the DH-PSF is performed with the calibration z-stack using ZOLA [59]. Localization precision curves displayed in Figure 6B have been smoothened using a rolling window of 15 points. Then, 3D superlocalization of diffusing uCCNTs in organotypic slices was performed in ZOLA using the phase-retrieved PSF. Trajectory reconstruction was then performed in Python using TrackPy library [60]. Trajectory handling and plotting is finally performed on custom Python routines.

**Data Processing and Analysis:**

**Brightness Analysis.** Bright analysis of the PL images at single nanotube level was performed using custom MATLAB (MathWorks) programs.

**Super-resolution Analysis of Color Center Density on ar- and ox-CCNTs.** The super-resolution analysis was performed on blinking ar- and ox-CCNTs when directly immobilized on a glass substrate. The $E_{11}$* blinking emission was monitored for 10000 frames and the sum of the intensities for each frame was fitted to a step-like signal, in a similar fashion as previously described by Cognet et al. [38], using the AutoStepFinder algorithm on Matlab [61]. The frames were then averaged for 10 frames before and after each step and subtracted to isolate each individual blinking event. A 2D Gaussian fitting was then applied to the subtracted frames in order to fit the position of the emitting sites. The 'X' and 'Y' positions of each emitting sites were determined along with their respective localization precisions. Those localizations showing a localization precision below 50 nm were then represented as the sum of 2D symmetrical Gaussians whose width is equal to the localization precision and center is equal to the X,Y localization of the emitter as shown in Figure 3D (Figure S12).

**Schematic Figures:**

Nanotube structures shown in Figure 1A and Figure 3E were created using the open-source Avogadro software.[62]

**SUPPLEMENTARY MATERIALS:** Supplementary materials include Figs. S1 to S16, Table S1 and Movie S1.

**NOTES**

The authors declare no competing financial interest.

**ACKNOWLEDGMENTS**

We would like to thank M. Tondusson for providing experimental help and support. L.C., L.G. and E.B. acknowledges financial support from the European Research Council Synergy grant (951294) and Agence Nationale de la Recherche (ANR-18-CE09-0019-02). L.C. acknowledges support from Agence Nationale de la Recherche (EUR Light&T, PIA3 Program, ANR-17-EURE-0027) and the Idex Bordeaux (Grand Research Program GPR LIGHT). S.N. and B.L. acknowledge funding from the European Union's Horizon 2020 research and innovation programme under the Marie Skłodowska-Curie grants agreement No. 101024294 and No. 101107105, respectively. F.L.S., S.S. and J.Z. acknowledge funding from the European Research Council (ERC) under the European Union's Horizon 2020 research and innovation programme (Grant Agreement No. 817494 "TRIFECTS"). J.Z. acknowledges additional support by the Deutsche Forschungsgemeinschaft (DFG, German Research Foundation) under Germany's Excellence Strategy for the Excellence Cluster "3D Matter Made to Order" (EXC−2082/1 – 390761711).

## REFERENCES

(1) Möckl, L.; Moerner, W. E. Super-Resolution Microscopy with Single Molecules in Biology and Beyond-Essentials, Current Trends, and Future Challenges. *Journal of the American Chemical Society* **2020**, *142* (42), 17828–17844. https://doi.org/10.1021/jacs.0c08178.

(2) Hell, S. W.; Sahl, S. J.; Bates, M.; Zhuang, X.; Heintzmann, R.; Booth, M. J.; Bewersdorf, J.; Shtengel, G.; Hess, H.; Tinnefeld, P.; Honigmann, A.; Jakobs, S.; Testa, I.; Cognet, L.; Lounis, B.; Ewers, H.; Davis, S. J.; Eggeling, C.; Klenerman, D.; Willig, K. I.; Vicidomini, G.; Castello, M.; Diaspro, A.; Cordes, T. The 2015 Super-Resolution Microscopy Roadmap. *Journal of Physics D: Applied Physics* **2015**, *48* (44), 443001. https://doi.org/10.1088/0022-3727/48/44/443001.

(3) Heine, M.; Groc, L.; Frischknecht, R.; Béïque, J.-C.; Lounis, B.; Rumbaugh, G.; Huganir, R. L.; Cognet, L.; Choquet, D. Surface Mobility of Postsynaptic AMPARs Tunes Synaptic Transmission. *Science* **2008**, *320* (5873), 201–205. https://doi.org/10.1051/medsci/2008245548.

(4) Liu, Z.; Lavis, L. D.; Betzig, E. Imaging Live-Cell Dynamics and Structure at the Single-Molecule Level. *Molecular Cell* **2015**, *58* (4), 644–659. https://doi.org/10.1016/j.molcel.2015.02.033.

(5) Tønnesen, J.; Inavalli, V. V. G. K.; Nägerl, U. V. Super-Resolution Imaging of the Extracellular Space in Living Brain Tissue. *Cell* **2018**, *172* (5), 1108-1121.e15. https://doi.org/10.1016/j.cell.2018.02.007.

(6) Jin, D.; Xi, P.; Wang, B.; Zhang, L.; Enderlein, J.; Van Oijen, A. M. Nanoparticles for Super-Resolution Microscopy and Single-Molecule Tracking. *Nature Methods* **2018**, *15* (6), 415–423. https://doi.org/10.1038/s41592-018-0012-4.

(7) Wang, L.; Frei, M. S.; Salim, A.; Johnsson, K. Small-Molecule Fluorescent Probes for Live-Cell Super-Resolution Microscopy. *Journal of the American Chemical Society* **2019**, *141* (7), 2770–2781. https://doi.org/10.1021/jacs.8b11134.

(8) Liu, S.; Hoess, P.; Ries, J. Super-Resolution Microscopy for Structural Cell Biology. *Annual Review of Biophysics* **2022**, *51* (Volume 51, 2022), 301–326. https://doi.org/10.1146/annurev-biophys-102521-112912.

(9) Chen, Y.; Wang, S.; Zhang, F. Near-Infrared Luminescence High-Contrast in Vivo Biomedical Imaging. *Nat Rev Bioeng* **2023**, *1* (1), 60–78. https://doi.org/10.1038/s44222-022-00002-8.

(10) Hong, G.; Antaris, A. L.; Dai, H. Near-Infrared Fluorophores for Biomedical Imaging. *Nature Biomedical Engineering* **2017**, *1* (1). https://doi.org/10.1038/s41551-016-0010.

(11) O'Connell, M. J.; Bachilo, S. M.; Huffman, C. B.; Moore, V. C.; Strano, M. S.; Haroz, E. H.; Rialon, K. L.; Boul, P. J.; Noon, W. H.; Kittrell, C.; Ma, J.; Hauge, R. H.; Weisman, R. B.; Smalley, R. E. Band Gap Fluorescence from Individual Single-Walled Carbon Nanotubes. *Science* **2002**, *297*, 593–596.

(12) Thimsen, E.; Sadtler, B.; Berezin, M. Y. Shortwave-Infrared (SWIR) Emitters for Biological Imaging: A Review of Challenges and Opportunities. *Nanophotonics* **2017**, *6* (5), 1043–1054. https://doi.org/10.1515/nanoph-2017-0039.

(13) Li, H.; Wang, X.; Ohulchanskyy, T. Y.; Chen, G. Lanthanide-Doped Near-Infrared Nanoparticles for Biophotonics. *Advanced Materials*. 2021, p 2000678.

(14) Godin, A. G.; Varela, J. A.; Gao, Z.; Danné, N.; Dupuis, J. P.; Lounis, B.; Groc, L.; Cognet, L. Single-Nanotube Tracking Reveals the Nanoscale Organization of the Extracellular Space in the Live Brain. *Nature Nanotechnology* **2017**, *12* (3), 238–243. https://doi.org/10.1038/nnano.2016.248.

(15) Soria, F. N.; Paviolo, C.; Doudnikoff, E.; Arotcarena, M. L.; Lee, A.; Danné, N.; Mandal, A. K.; Gosset, P.; Dehay, B.; Groc, L.; Cognet, L.; Bezard, E. Synucleinopathy Alters Nanoscale Organization and Diffusion in the Brain Extracellular Space through Hyaluronan Remodeling. *Nature Communications* **2020**, *11* (1). https://doi.org/10.1038/s41467-020-17328-9.

(16) Kruss, S.; Salem, D. P.; Vukovic, L.; Lima, B.; Vander Ende, E.; Boyden, E. S.; Strano, M. S. High-Resolution Imaging of Cellular Dopamine Efflux Using a Fluorescent Nanosensor Array. *PNAS* **2017**, *114* (8), 1789–1794. https://doi.org/10.1073/pnas.1613541114.

(17) Galassi, T. V.; Jena, P. V.; Shah, J.; Ao, G.; Molitor, E.; Bram, Y.; Frankel, A.; Park, J.; Jessurun, J.; Ory, D. S.; Haimovitz-Friedman, A.; Roxbury, D.; Mittal, J.; Zheng, M.; Schwartz, R. E.; Heller, D. A. An Optical Nanoreporter of Endolysosomal Lipid Accumulation Reveals Enduring Effects of Diet on Hepatic Macrophages in Vivo. *Science Translational Medicine* **2018**, *10* (461), 1–11. https://doi.org/10.1126/scitranslmed.aar2680.

(18) Bulumulla, C.; Krasley, A. T.; Cristofori-Armstrong, B.; Valinsky, W. C.; Walpita, D.; Ackerman, D.; Clapham, D. E.; Beyene, A. G. Visualizing Synaptic Dopamine Efflux with a 2D Composite Nanofilm. *eLife* **2022**, *11*, 1–27. https://doi.org/10.7554/elife.78773.

(19) Cognet, L.; Tsyboulski, D. A.; Rocha, J.-D. R.; Doyle, C. D.; Tour, J. M.; Weisman, R. B. Stepwise Quenching of Exciton Fluorescence in Carbon Nanotubes by Single-Molecule Reactions. *Science* **2007**, *316* (5830), 1465–1468. https://doi.org/10.1126/science.1141316.

(20) Fakhri, N.; MacKintosh, F. C.; Lounis, B.; Cognet, L.; Pasquali, M. Brownian Motion of Stiff Filaments in a Crowded Environment. *Science* **2010**, *330* (6012), 1804–1807. https://doi.org/10.1126/science.1197321.

(21) Bacanu, A.; Pelletier, J. F.; Jung, Y.; Fakhri, N. Inferring Scale-Dependent Non-Equilibrium Activity Using Carbon Nanotubes. *Nat. Nanotechnol.* **2023**, *18* (8), 905–911. https://doi.org/10.1038/s41565-023-01395-2.

(22) Piao, Y.; Meany, B.; Powell, L. R.; Valley, N.; Kwon, H.; Schatz, G. C.; Wang, Y. Brightening of Carbon Nanotube Photoluminescence through the Incorporation of Sp 3 Defects. *Nature Chemistry* **2013**, *5* (10), 840–845. https://doi.org/10.1038/nchem.1711.

(23) Zaumseil, J. Luminescent Defects in Single-Walled Carbon Nanotubes for Applications. *Advanced Optical Materials* **2022**.

(24) Ghosh, S.; Bachilo, S. M.; Simonette, R. A.; Beckingham, K. M.; Weisman, R. B. Oxygen Doping Modifies Near-Infrared Band Gaps in Fluorescent Single-Walled Carbon Nanotubes. *Science* **2010**, *330* (6011), 1656–1659. https://doi.org/10.1126/science.1196382.

(25) Danné, N.; Kim, M.; Godin, A. G.; Kwon, H.; Gao, Z.; Wu, X.; Hartmann, N. F.; Doorn, S. K.; Lounis, B.; Wang, Y.; Cognet, L. Ultrashort Carbon Nanotubes That Fluoresce Brightly in the Near-Infrared. *ACS Nano* **2018**, *12* (6), 6059–6065. https://doi.org/10.1021/acsnano.8b02307.

(26) Li, Y.; Wu, X.; Kim, M.; Fortner, J.; Qu, H.; Wang, Y. Fluorescent Ultrashort Nanotubes from Defect-Induced Chemical Cutting. *Chemistry of Materials* **2019**, *31* (12), 4536–4544. https://doi.org/10.1021/acs.chemmater.9b01196.

(27) Welsher, K.; Liu, Z.; Sherlock, S. P.; Robinson, J. T.; Chen, Z.; Daranciang, D.; Dai, H. A Route to Brightly Fluorescent Carbon Nanotubes for Near-Infrared Imaging in Mice. *Nature Nanotech* **2009**, *4* (11), 773–780. https://doi.org/10.1038/nnano.2009.294.
(28) Cherukuri, T. K.; Tsyboulski, D. A.; Weisman, R. B. Length- and Defect-Dependent Fluorescence Efficiencies of Individual Single-Walled Carbon Nanotubes. *ACS Nano* **2012**, *6* (1), 843–850. https://doi.org/10.1021/nn2043516.
(29) Settele, S.; Stammer, F.; Sebastian, F. L.; Lindenthal, S.; Wald, S. R.; Li, H.; Flavel, B. S.; Zaumseil, J. Easy Access to Bright Oxygen Defects in Biocompatible Single-Walled Carbon Nanotubes via a Fenton-like Reaction. *ACS Nano* **2024**. https://doi.org/10.1021/acsnano.4c06448.
(30) Haggenmueller, R.; Rahatekar, S. S.; Fagan, J. A.; Chun, O. J.; Becker, M. L.; Naik, O. R. R.; Krauss, T.; Carlson, L.; Kadla, J. F.; Trulove, P. C.; Fox, D. F.; Delong, H. C.; Fang, Z.; Kelley, S. O.; Gilman, J. W.; Base, W. A. F.; Uni, V.; York, N.; Chemistry, B.; Vancou, V.; Vt, B. C.; Academy, U. S. N. V; Science, W. A.; Chun, J.; Becker, M. L.; Naik, R. R.; Krauss, T.; Carlson, L.; Kadla, J. F.; Trulove, P. C.; Fox, D. F.; Delong, H. C.; Fang, Z.; Kelley, S. O.; Gilrnan, J. W.; Chun, O. J.; Becker, M. L.; Naik, O. R. R.; Krauss, T.; Carlson, L.; Kadla, J. F.; Trulove, P. C.; Fox, D. F.; Delong, H. C.; Fang, Z.; Kelley, S. O.; Gilman, J. W.; Base, W. A. F.; Uni, V.; York, N.; Chemistry, B.; Vancou, V.; Vt, B. C.; Academy, U. S. N. V; Science, W. A. Comparison of the Quality of Aqueous Dispersions of Single Wall Carbon Nanotubes Using Surfactants and Biomolecules. *Langmuir* **2008**, *24* (10), 5070–5078. https://doi.org/10.1021/la703008r.
(31) Dong, L.; Witkowski, C. M.; Craig, M. M.; Greenwade, M. M.; Joseph, K. L. Cytotoxicity Effects of Different Surfactant Molecules Conjugated to Carbon Nanotubes on Human Astrocytoma Cells. *Nanoscale Research Letters* **2009**, *4* (12), 1517–1523. https://doi.org/10.1007/s11671-009-9429-0.
(32) Gao, Z.; Danné, N.; Godin, A.; Lounis, B.; Cognet, L. Evaluation of Different Single-Walled Carbon Nanotube Surface Coatings for Single-Particle Tracking Applications in Biological Environments. *Nanomaterials* **2017**, *7* (11), 393. https://doi.org/10.3390/nano7110393.
(33) Iwamura, M.; Akizuki, N.; Miyauchi, Y.; Mouri, S.; Shaver, J.; Gao, Z.; Cognet, L.; Lounis, B.; Matsuda, K. Nonlinear Photoluminescence Spectroscopy of Carbon Nanotubes with Localized Exciton States. *ACS Nano* **2014**, *8* (11), 11254–11260. https://doi.org/10.1021/nn503803b.
(34) Mandal, A. K.; Wu, X.; Ferreira, J. S.; Kim, M.; Powell, L. R.; Kwon, H.; Groc, L.; Wang, Y. H.; Cognet, L. Fluorescent Sp3 Defect-Tailored Carbon Nanotubes Enable NIR-II Single Particle Imaging in Live Brain Slices at Ultra-Low Excitation Doses. *Scientific Reports* **2020**, *10* (1), 1–9. https://doi.org/10.1038/s41598-020-62201-w.
(35) Streit, J. K.; Bachilo, S. M.; Ghosh, S.; Lin, C.-W.; Weisman, R. B. Directly Measured Optical Absorption Cross Sections for Structure-Selected Single-Walled Carbon Nanotubes. *Nano Lett.* **2014**, *14* (3), 1530–1536. https://doi.org/10.1021/nl404791y.
(36) Ashoka, A. H.; Aparin, I. O.; Reisch, A.; Klymchenko, A. S. Brightness of Fluorescent Organic Nanomaterials. *Chem. Soc. Rev.* **2023**, *52* (14), 4525–4548. https://doi.org/10.1039/D2CS00464J.
(37) Miyauchi, Y.; Iwamura, M.; Mouri, S.; Kawazoe, T.; Ohtsu, M.; Matsuda, K. Brightening of Excitons in Carbon Nanotubes on Dimensionality Modification. *Nature Photonics* **2013**, *7* (9), 715–719. https://doi.org/10.1038/nphoton.2013.179.

(38) Cognet, L.; Tsyboulski, D. A.; Weisman, R. B. Subdiffraction Far-Field Imaging of Luminescent Single-Walled Carbon Nanotubes. *Nano Letters* **2008**, *8* (2), 749–753. https://doi.org/10.1021/nl0725300.
(39) Ishii, A.; Yoshida, M.; Kato, Y. K. Exciton Diffusion, End Quenching, and Exciton-Exciton Annihilation in Individual Air-Suspended Carbon Nanotubes. *Phys. Rev. B* **2015**, *91* (12), 125427. https://doi.org/10.1103/PhysRevB.91.125427.
(40) Ma, X.; Adamska, L.; Yamaguchi, H.; Yalcin, S. E.; Tretiak, S.; Doorn, S. K.; Htoon, H. Electronic Structure and Chemical Nature of Oxygen Dopant States in Carbon Nanotubes. *ACS Nano* **2014**, *8* (10), 10782–10789. https://doi.org/10.1021/nn504553y.
(41) Sebastian, F. L.; Zorn, N. F.; Settele, S.; Lindenthal, S.; Berger, F. J.; Bendel, C.; Li, H.; Flavel, B. S.; Zaumseil, J. Absolute Quantification of Sp $^{3}$ Defects in Semiconducting Single-Wall Carbon Nanotubes by Raman Spectroscopy. *J. Phys. Chem. Lett.* **2022**, *13* (16), 3542–3548. https://doi.org/10.1021/acs.jpclett.2c00758.
(42) Sebastian, F. L.; Settele, S.; Li, H.; Flavel, B. S.; Zaumseil, J. How to Recognize Clustering of Luminescent Defects in Single-Wall Carbon Nanotubes. *Nanoscale Horiz.* **2024**. https://doi.org/10.1039/D4NH00383G.
(43) Lefebvre, J.; Austing, D. G.; Bond, J.; Finnie, P. Photoluminescence Imaging of Suspended Single-Walled Carbon Nanotubes. *Nano Lett.* **2006**, *6* (8), 1603–1608. https://doi.org/10.1021/nl060530e.
(44) Carlson, L. J.; Maccagnano, S. E.; Zheng, M.; Silcox, J.; Krauss, T. D. Fluorescence Efficiency of Individual Carbon Nanotubes. *Nano Lett.* **2007**, *7* (12), 3698–3703. https://doi.org/10.1021/nl072014+.
(45) Berciaud, S.; Cognet, L.; Lounis, B. Luminescence Decay and the Absorption Cross Section of Individual Single-Walled Carbon Nanotubes. *Phys. Rev. Lett.* **2008**, *101* (7), 077402. https://doi.org/10.1103/PhysRevLett.101.077402.
(46) Sarkar, S.; Le, P.; Geng, J.; Liu, Y.; Han, Z.; Zahid, M. U.; Nall, D.; Youn, Y.; Selvin, P. R.; Smith, A. M. Short-Wave Infrared Quantum Dots with Compact Sizes as Molecular Probes for Fluorescence Microscopy. *J. Am. Chem. Soc.* **2020**, *142* (7), 3449–3462. https://doi.org/10.1021/jacs.9b11567.
(47) Kubicek-Sutherland, J. Z.; Makarov, N. S.; Stromberg, Z. R.; Lenz, K. D.; Castañeda, C.; Mercer, A. N.; Mukundan, H.; McDaniel, H.; Ramasamy, K. Exploring the Biocompatibility of Near-IR CuInSexS2–x/ZnS Quantum Dots for Deep-Tissue Bioimaging. *ACS Appl. Bio Mater.* **2020**, *3* (12), 8567–8574. https://doi.org/10.1021/acsabm.0c00939.
(48) Zhang, Y.; Hong, G.; Zhang, Y.; Chen, G.; Li, F.; Dai, H.; Wang, Q. Ag2S Quantum Dot: A Bright and Biocompatible Fluorescent Nanoprobe in the Second Near-Infrared Window. *ACS Nano* **2012**, *6* (5), 3695–3702. https://doi.org/10.1021/nn301218z.
(49) Zhang, M.; Yue, J.; Cui, R.; Ma, Z.; Wan, H.; Wang, F.; Zhu, S.; Zhou, Y.; Kuang, Y.; Zhong, Y.; Pang, D.-W.; Dai, H. Bright Quantum Dots Emitting at ∼1,600 Nm in the NIR-IIb Window for Deep Tissue Fluorescence Imaging. *Proc. Natl. Acad. Sci. U.S.A.* **2018**, *115* (26), 6590–6595. https://doi.org/10.1073/pnas.1806153115.
(50) Hrabetova, S.; Cognet, L.; Rusakov, D. A.; Nägerl, U. V. Unveiling the Extracellular Space of the Brain: From Super-Resolved Microstructure to *In Vivo* Function. *J. Neurosci.* **2018**, *38* (44), 9355–9363. https://doi.org/10.1523/JNEUROSCI.1664-18.2018.
(51) Tønnesen, J.; Hrabětová, S.; Soria, F. N. Local Diffusion in the Extracellular Space of the Brain. *Neurobiology of Disease* **2023**, *177*, 105981. https://doi.org/10.1016/j.nbd.2022.105981.

(52) Lelek, M.; Gyparaki, M. T.; Beliu, G.; Schueder, F.; Griffié, J.; Manley, S.; Jungmann, R.; Sauer, M.; Lakadamyali, M.; Zimmer, C. Single-Molecule Localization Microscopy. *Nat Rev Methods Primers* **2021**, *1* (1), 1–27. https://doi.org/10.1038/s43586-021-00038-x.

(53) Pavani, S. R. P.; Thompson, M. A.; Biteen, J. S.; Lord, S. J.; Liu, N.; Twieg, R. J.; Piestun, R.; Moerner, W. E. Three-Dimensional, Single-Molecule Fluorescence Imaging beyond the Diffraction Limit by Using a Double-Helix Point Spread Function. *Proceedings of the National Academy of Sciences of the United States of America* **2009**, *106* (9), 2995–2999. https://doi.org/10.1073/pnas.0900245106.

(54) Li, H.; Gordeev, G.; Garrity, O.; Reich, S.; Flavel, B. S. Separation of Small-Diameter Single-Walled Carbon Nanotubes in One to Three Steps with Aqueous Two-Phase Extraction. *ACS Nano* **2019**, *13* (2), 2567–2578. https://doi.org/10.1021/acsnano.8b09579.

(55) Graf, A.; Zakharko, Y.; Schießl, S. P.; Backes, C.; Pfohl, M.; Flavel, B. S.; Zaumseil, J. Large Scale, Selective Dispersion of Long Single-Walled Carbon Nanotubes with High Photoluminescence Quantum Yield by Shear Force Mixing. *Carbon* **2016**, *105*, 593–599. https://doi.org/10.1016/j.carbon.2016.05.002.

(56) Pfohl, M.; Tune, D. D.; Graf, A.; Zaumseil, J.; Krupke, R.; Flavel, B. S. Fitting Single-Walled Carbon Nanotube Optical Spectra. *ACS Omega* **2017**, *2* (3), 1163–1171. https://doi.org/10.1021/acsomega.6b00468.

(57) Ovesný, M.; Křížek, P.; Borkovec, J.; Švindrych, Z.; Hagen, G. M. ThunderSTORM: A Comprehensive ImageJ Plug-in for PALM and STORM Data Analysis and Super-Resolution Imaging. *Bioinformatics* **2014**, *30* (16), 2389–2390. https://doi.org/10.1093/bioinformatics/btu202.

(58) Paviolo, C.; Ferreira, J. S.; Lee, A.; Hunter, D.; Calaresu, I.; Nandi, S.; Groc, L.; Cognet, L. Near-Infrared Carbon Nanotube Tracking Reveals the Nanoscale Extracellular Space around Synapses. *Nano Lett.* **2022**, *22* (17), 6849–6856. https://doi.org/10.1021/acs.nanolett.1c04259.

(59) Aristov, A.; Lelandais, B.; Rensen, E.; Zimmer, C. ZOLA-3D Allows Flexible 3D Localization Microscopy over an Adjustable Axial Range. *Nat Commun* **2018**, *9* (1), 2409. https://doi.org/10.1038/s41467-018-04709-4.

(60) Allan, D. B.; Caswell, T.; Keim, N. C.; van der Wel, C. M.; Verweij, R. W. *Soft-Matter/Trackpy,* V0.6.4; Zenodo, **2024**. https://doi.org/10.5281/zenodo.12708864.

(61) Loeff, L.; Kerssemakers, J. W. J.; Joo, C.; Dekker, C. AutoStepfinder: A Fast and Automated Step Detection Method for Single-Molecule Analysis. *Patterns* **2021**, *2* (5), 100256. https://doi.org/10.1016/j.patter.2021.100256.

(62) Hanwell, M. D.; Curtis, D. E.; Lonie, D. C.; Vandermeersch, T.; Zurek, E.; Hutchison, G. R. Avogadro: An Advanced Semantic Chemical Editor, Visualization, and Analysis Platform. *J Cheminform* **2012**, *4* (1), 1–17. https://doi.org/10.1186/1758-2946-4-17.

SUPPLEMENTARY MATERIAL

# Ultrashort Carbon Nanotubes with Luminescent Color Centers are Bright NIR-II Nano-Emitters

Somen Nandi[1,2]†, Quentin Gresil[1,2]†, Benjamin P. Lambert[1,2], Finn L. Sebastian[3], Simon Settele[3], Ivo Calaresu[4], Juan Estaun-Panzano[5], Anna Lovisotto[5], Claire Mazzocco[5], Benjamin S. Flavel[6], Erwan Bezard[5], Laurent Groc[4], Jana Zaumseil[3], Laurent Cognet[1,2,*]

[1]Laboratoire Photonique Numérique et Nanosciences, Université de Bordeaux, 33400 Talence, France

[2]LP2N, Institut d'Optique Graduate School, CNRS UMR 5298, 33400 Talence, France

[3]Institute for Physical Chemistry, Heidelberg University, D-69120 Heidelberg, Germany

[4]Interdisciplinary Institute for Neuroscience, CNRS, Univ. Bordeaux, 33076 Bordeaux, France

[5]Univ. Bordeaux, CNRS, IMN, UMR 5293, F-33000 Bordeaux, France

[6]Institute of Nanotechnology, Karlsruhe Institute of Technology, Kaiserstraße 12, D-76131 Karlsruhe, Germany

*Correspondence: laurent.cognet@u-bordeaux.fr

†These authors contributed equally to this work.

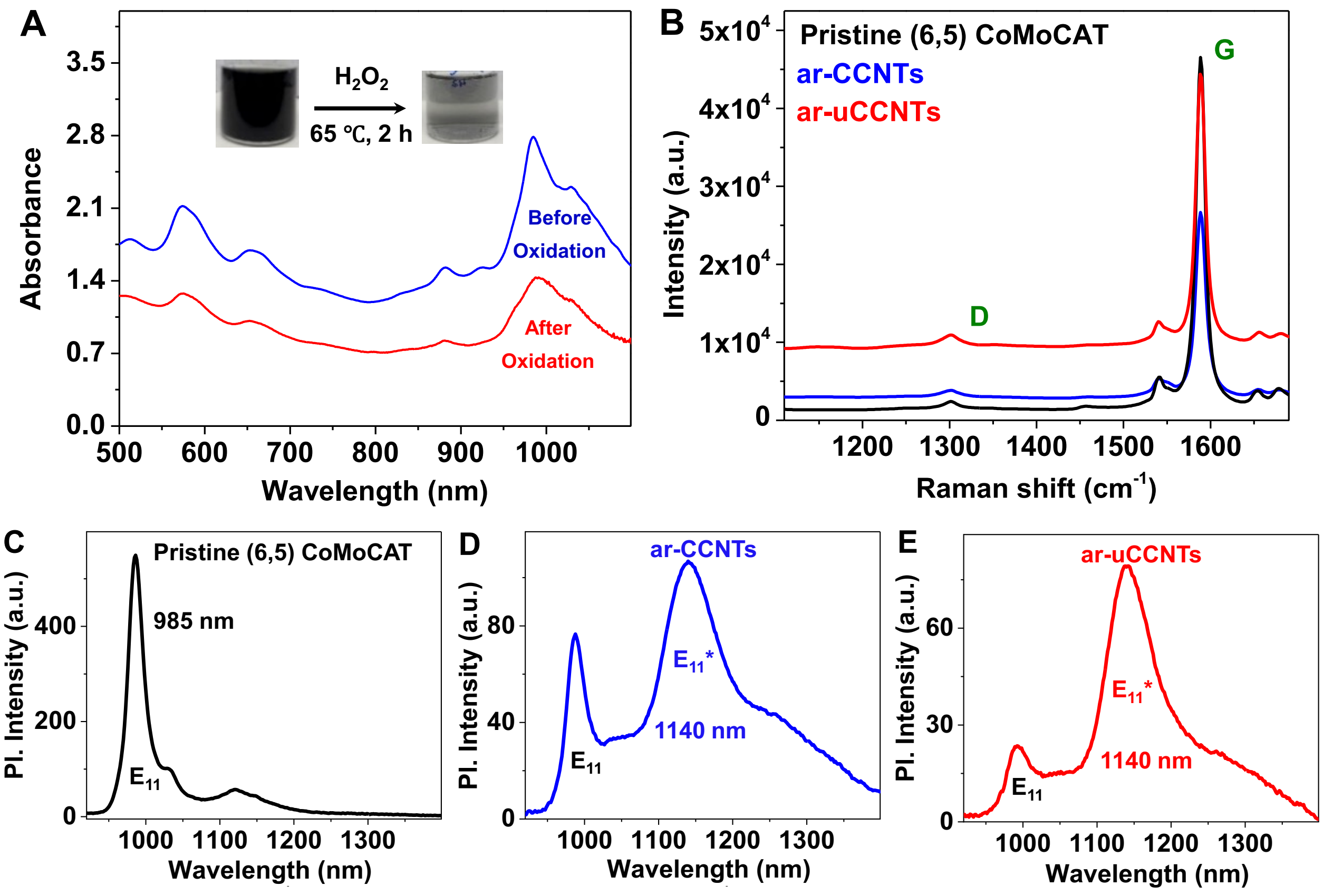

**Figure S1**. (A) UV-vis-NIR absorption spectra of the ar-CCNTs before (blue) and after (red) $H_2O_2$ oxidation in 1% DOC. Inset shows the images of the dispersions before and after the $H_2O_2$ cutting process. (B) Raman spectra (excitation line is 638 nm) of pristine (6,5) CoMoCAT (black), ar-CCNTs (blue) and ar-uCCNTs (red); (C-E) Corresponding their PL spectra (excitation wavelength is 568 nm).

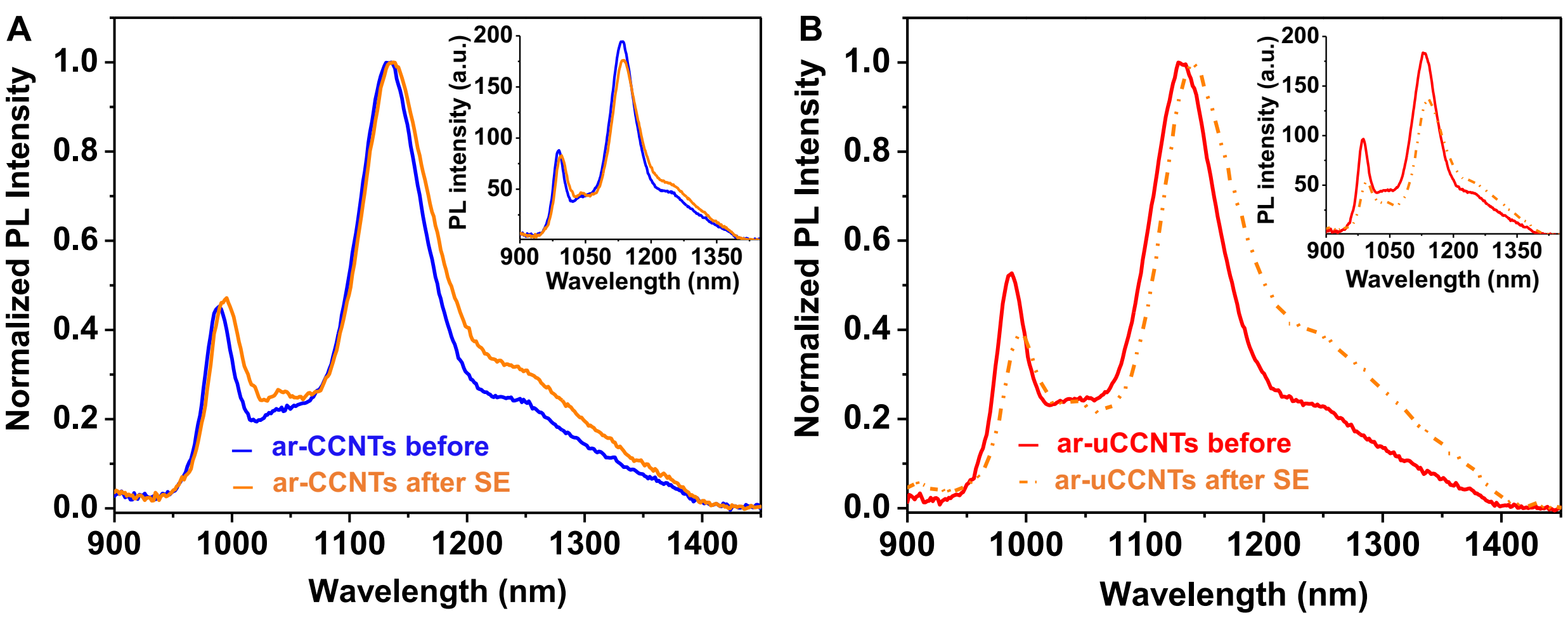

**Figure S2**. **PL spectra of surfactant exchanged (SE) samples by PLPEG**. (A) ar-CCNTs and (B) ar-uCCNTs. Corresponding un-normalized spectra are shown in the inset

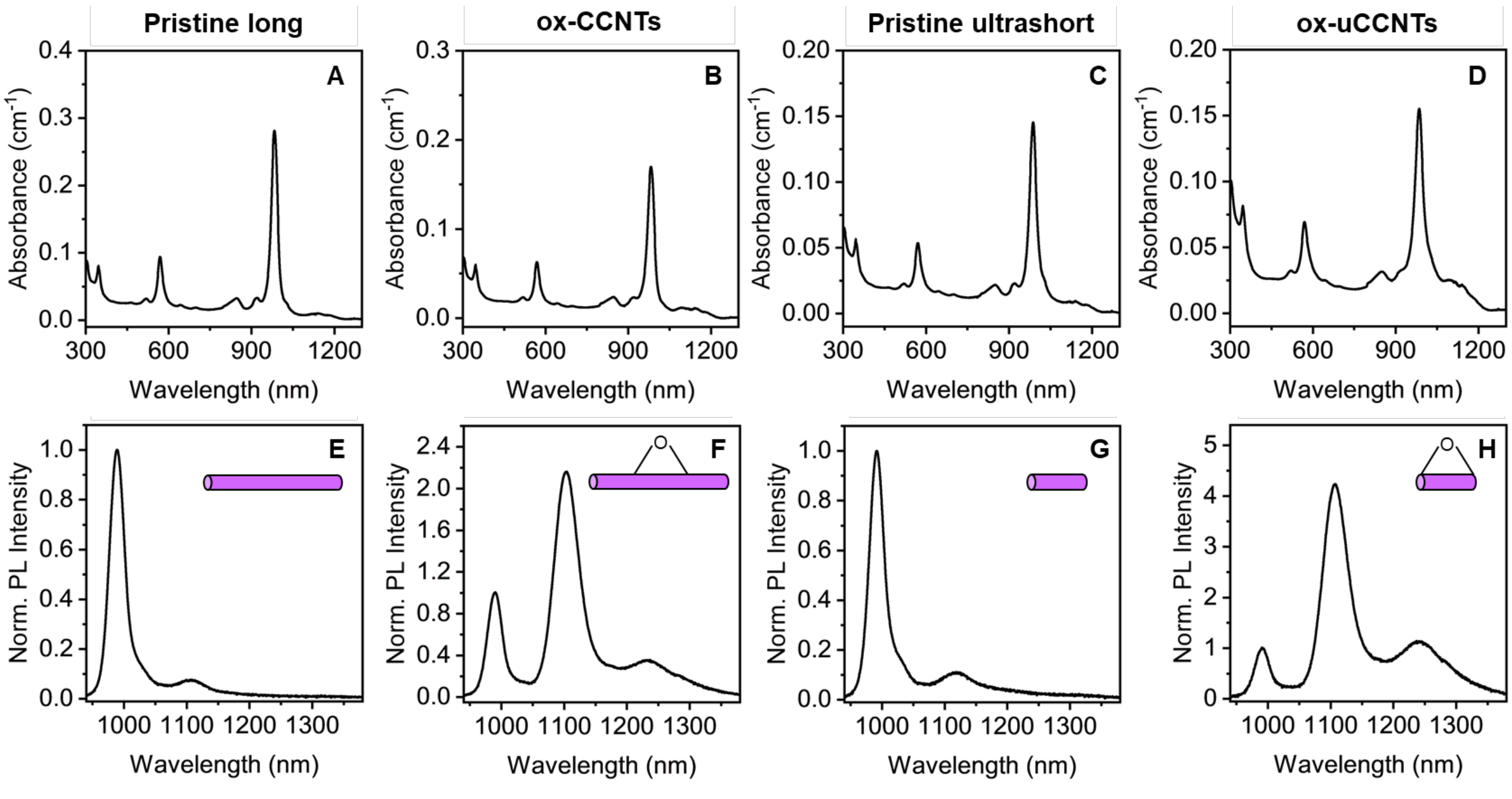

**Figure S3**. (A-D) UV-vis-NIR absorption spectra of pristine long (A) and ultrashort (C) SWCNTs along with the oxygen functionalized samples, (B) and (D), respectively, in 1% DOC. (E-H) Corresponding PL spectra.

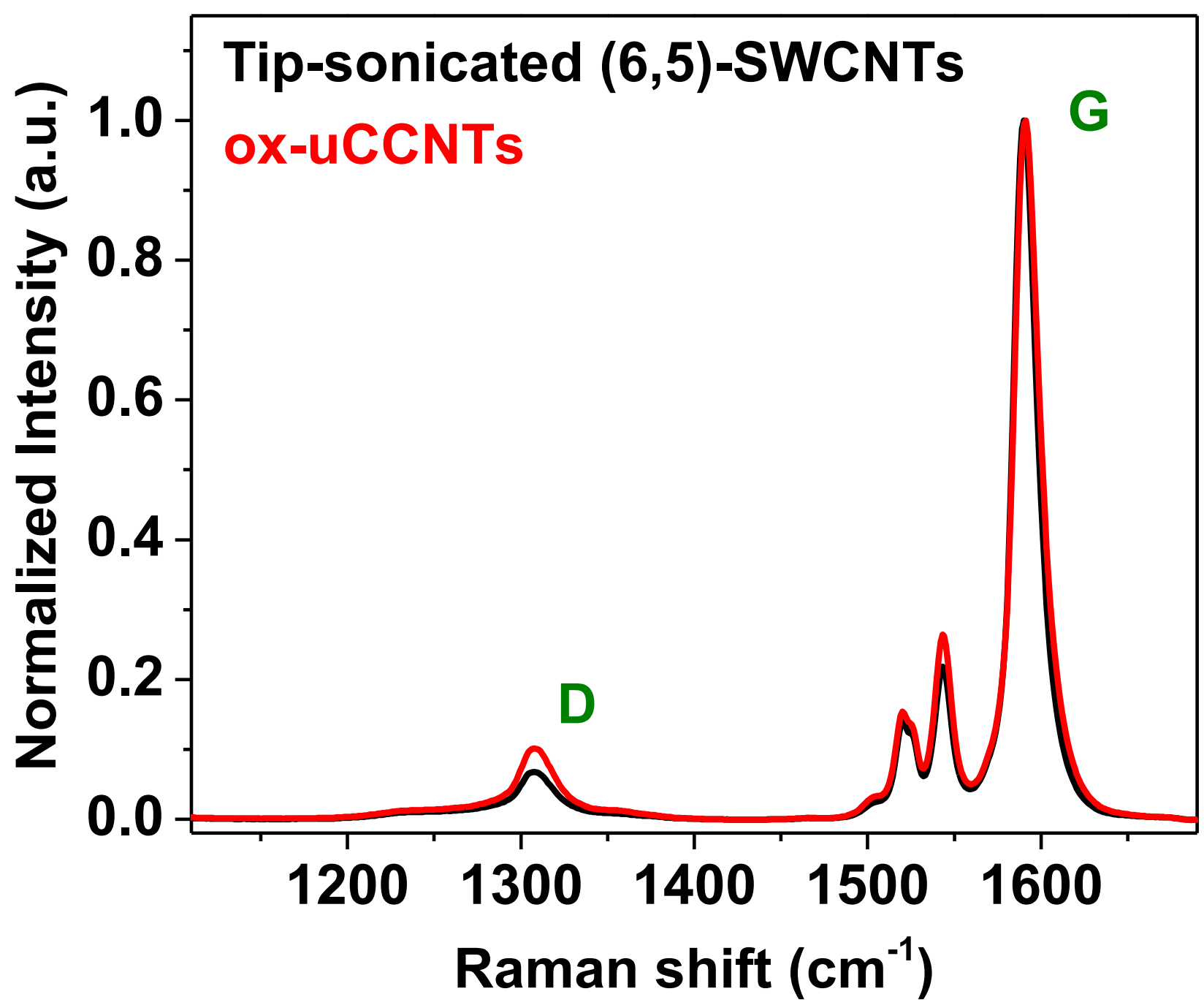

**Figure S4**. Raman spectra (excitation line is 532 nm) of tip-sonicated, unfunctionalized (6,5)-SWCNTs (black) and ox-uCCNTs (red).

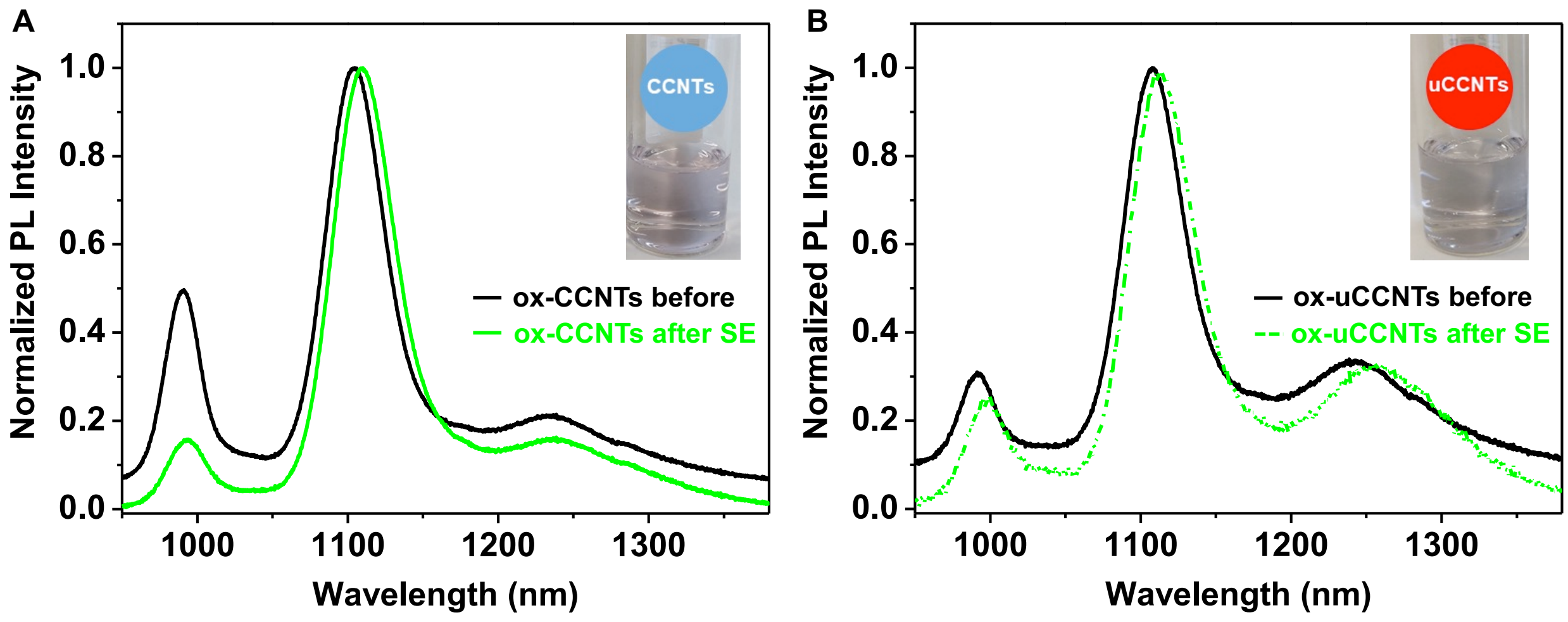

**Figure S5. PL spectra of surfactant exchanged (SE) samples by PLPEG**. (A) ox-CCNTs and (B) ox-uCCNTs. Inset shows the images of their corresponding dispersion after surfactant exchange to PLPEG.

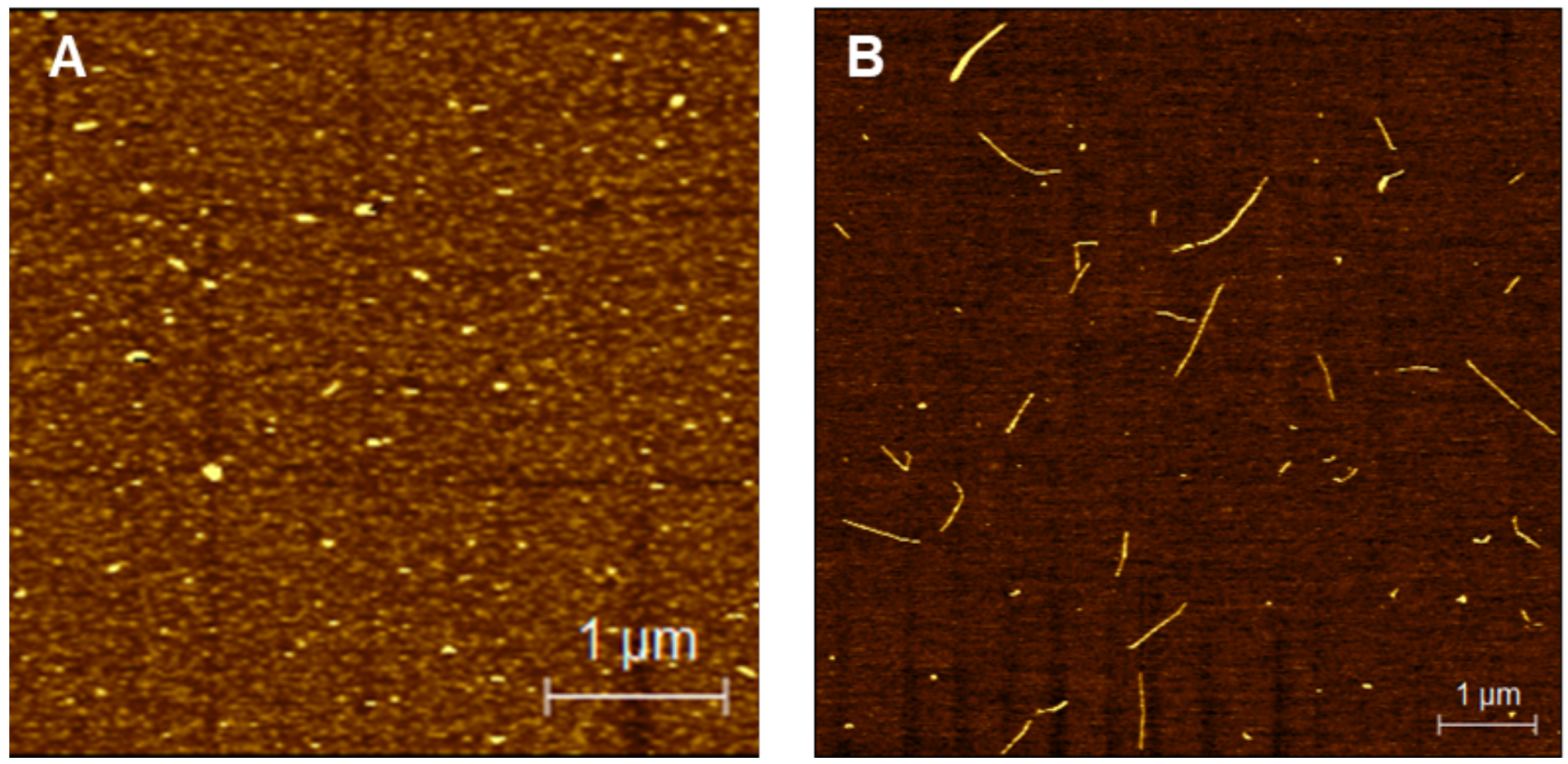

**Figure S6**. AFM images of (A) ar-uCCNTs and (B) ar-CCNTs, drop casted on mica surface and recorded with an AFM from Oxford Instruments, MFP-3D Infinity, Asylum research.

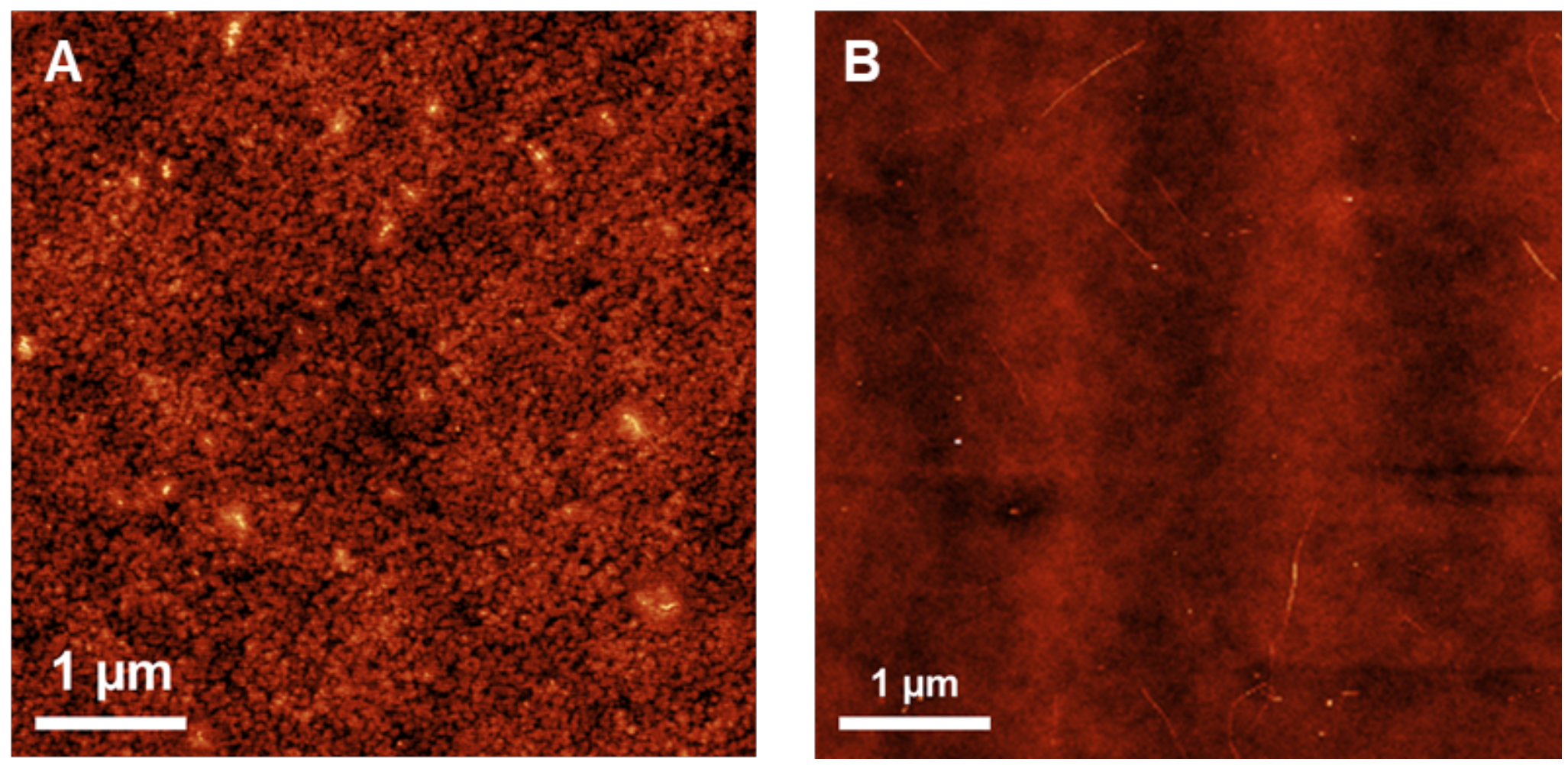

**Figure S7**. AFM images of (A) ox-uCCNTs and (B) ox-CCNTs, drop casted on $Si/SiO_2$ substrates and recorded with a Bruker Dimension Icon AFM.

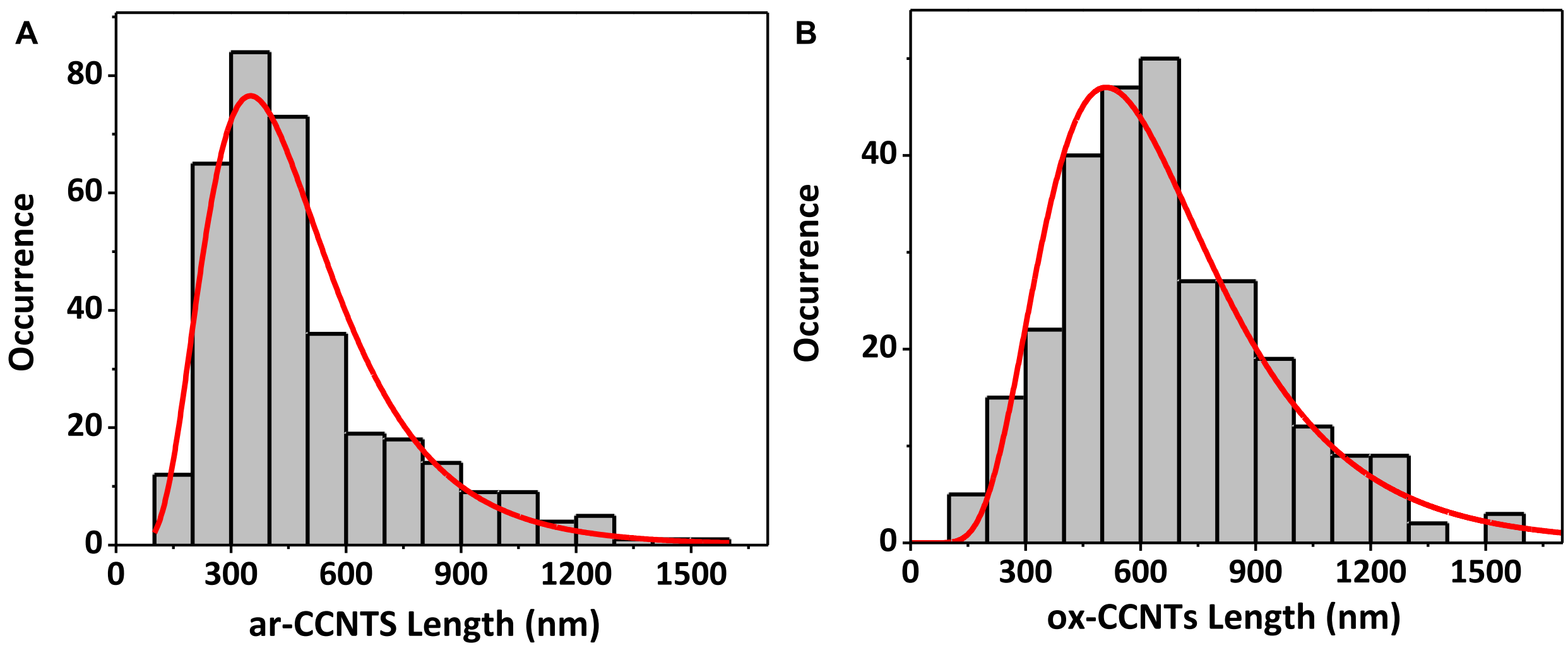

**Figure S8**. AFM length distributions of (A) ar-CCNTs (Median length = 410 nm, Mean length = 490 ± 250 nm) and (B) ox-CCNTs (Median length = 626 nm, Mean length = 674 ± 280 nm), where more than 275 CCNTs are measured for each sample.

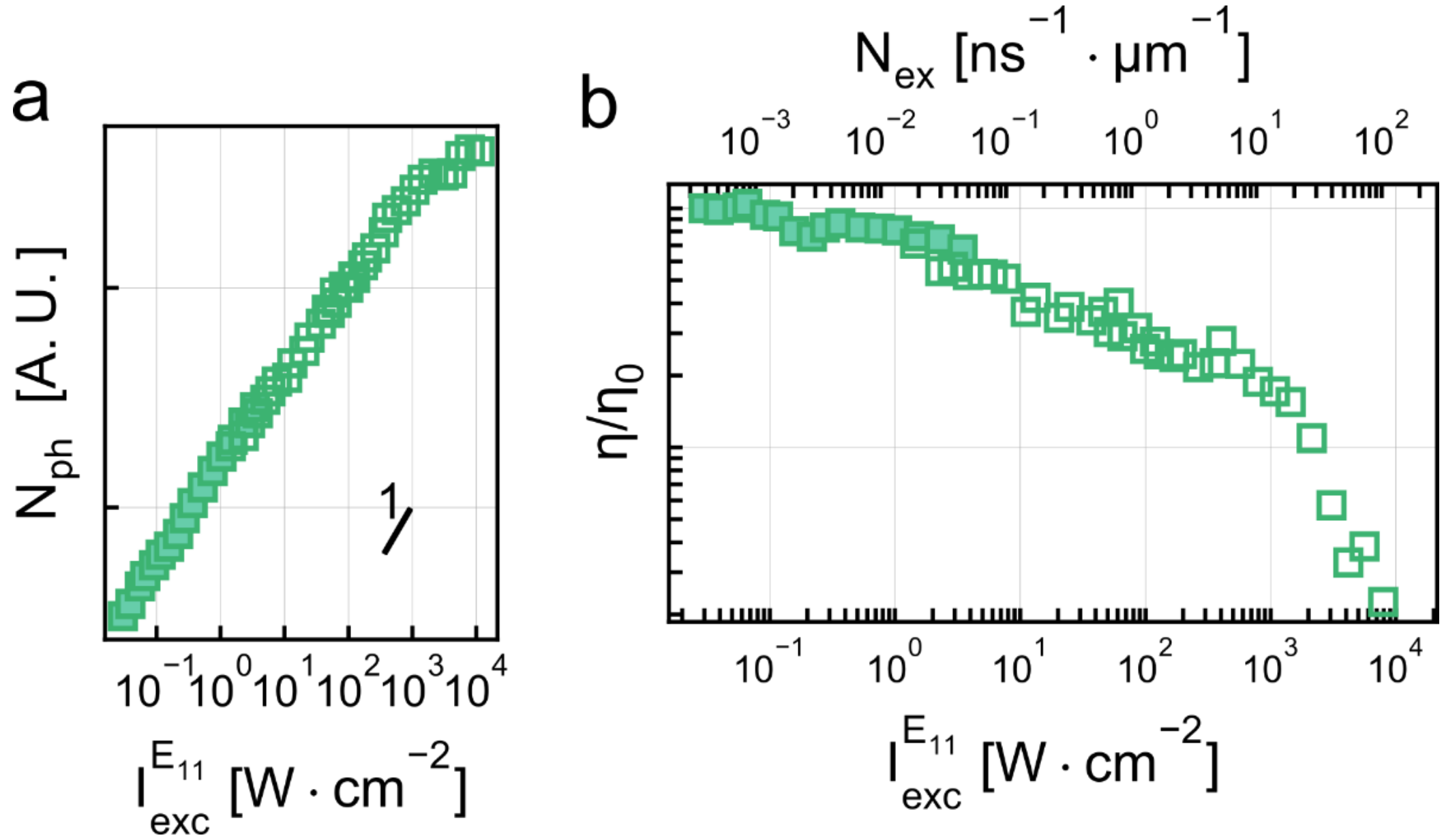

**Figure S9**: PL Dependency on Excitation Intensity for ox-uCCNTs. (A) $E_{11}$* PL intensities as functions of cw excitation intensity ($E_{11}$ excitation), $I_{exc}^{E_{11}}$, obtained from ensemble (filled symbols) and single nanotube measurements (open symbols). (B) The dependence of normalized PLQY $(\eta/\eta_0)$ on $I_{exc}^{E_{11}}$ is obtained by dividing the measured PL values by $I_{exc}^{E_{11}}$, The corresponding exciton density $N_{exc}$ (in µm$^{-1}$·ns$^{-1}$) generated in nanotubes upon light absorption is shown on the top x-axis.

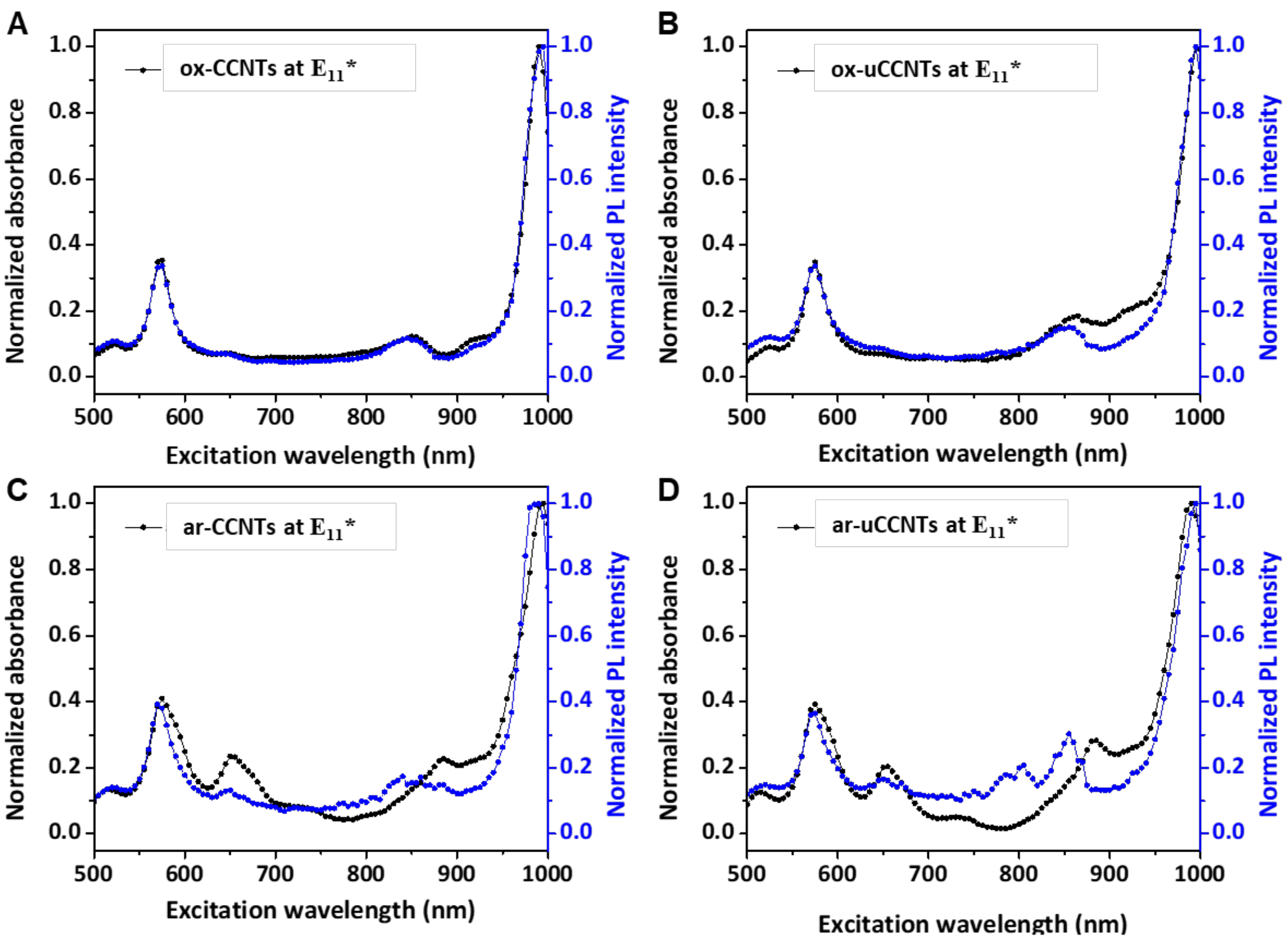

**Figure S10**. Comparison of absorption spectra of the nanotube dispersions with their PL excitation (PLE) spectra by monitoring the $E_{11}$* PL intensity.

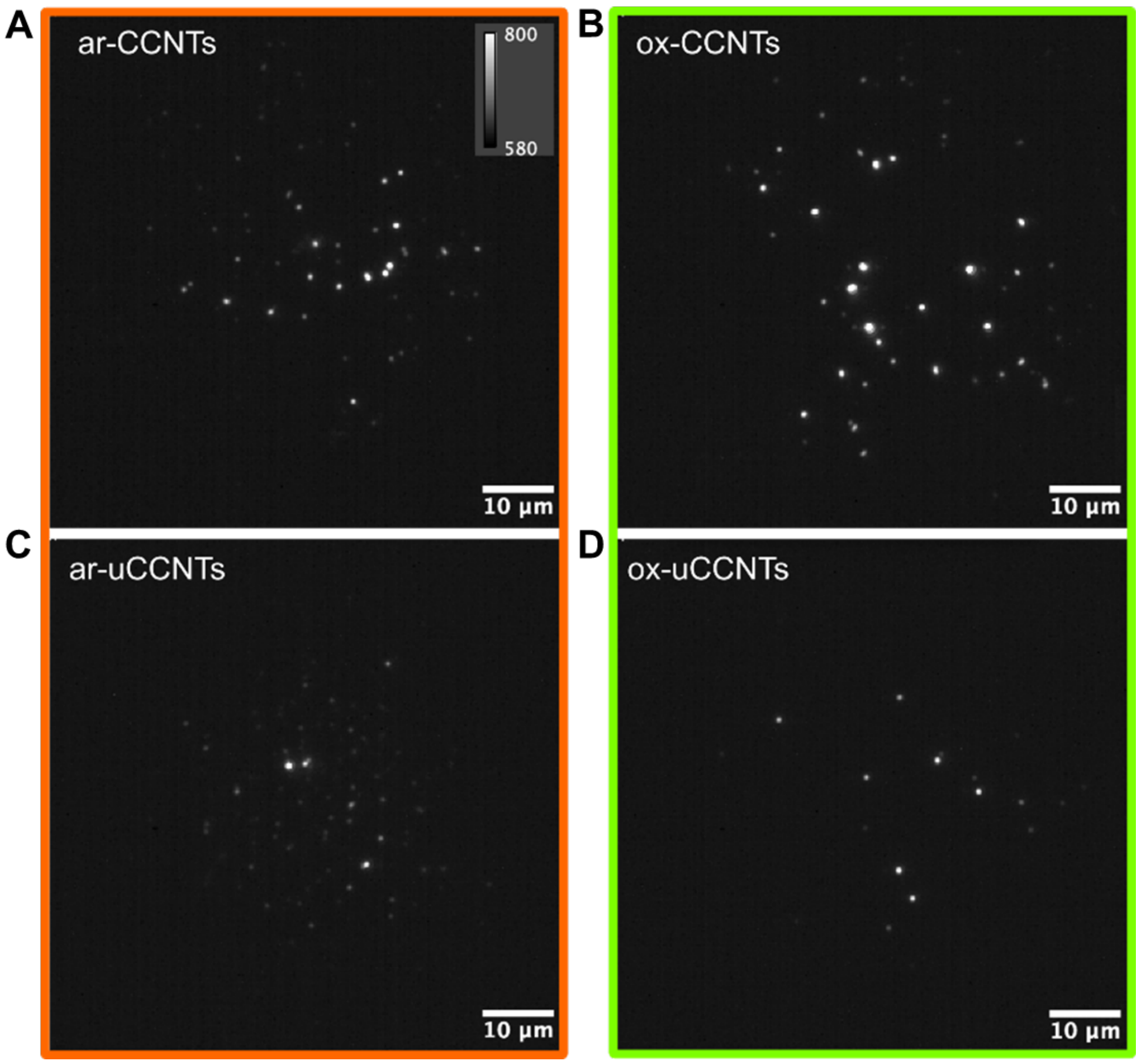

**Figure S11**. **Single molecule microscopy imaging.** Images of $E_{11}$* PL from individual nanotubes recorded using a 985 nm excitation and 10 ms exposure time; excitation intensity: 200 W/cm$^2$: (A) ar-CCNTs, (B) ox-CCNTs, (C) ar-uCCNTs and (D) ox-uCCNTs. Scale bars: 10 µm.

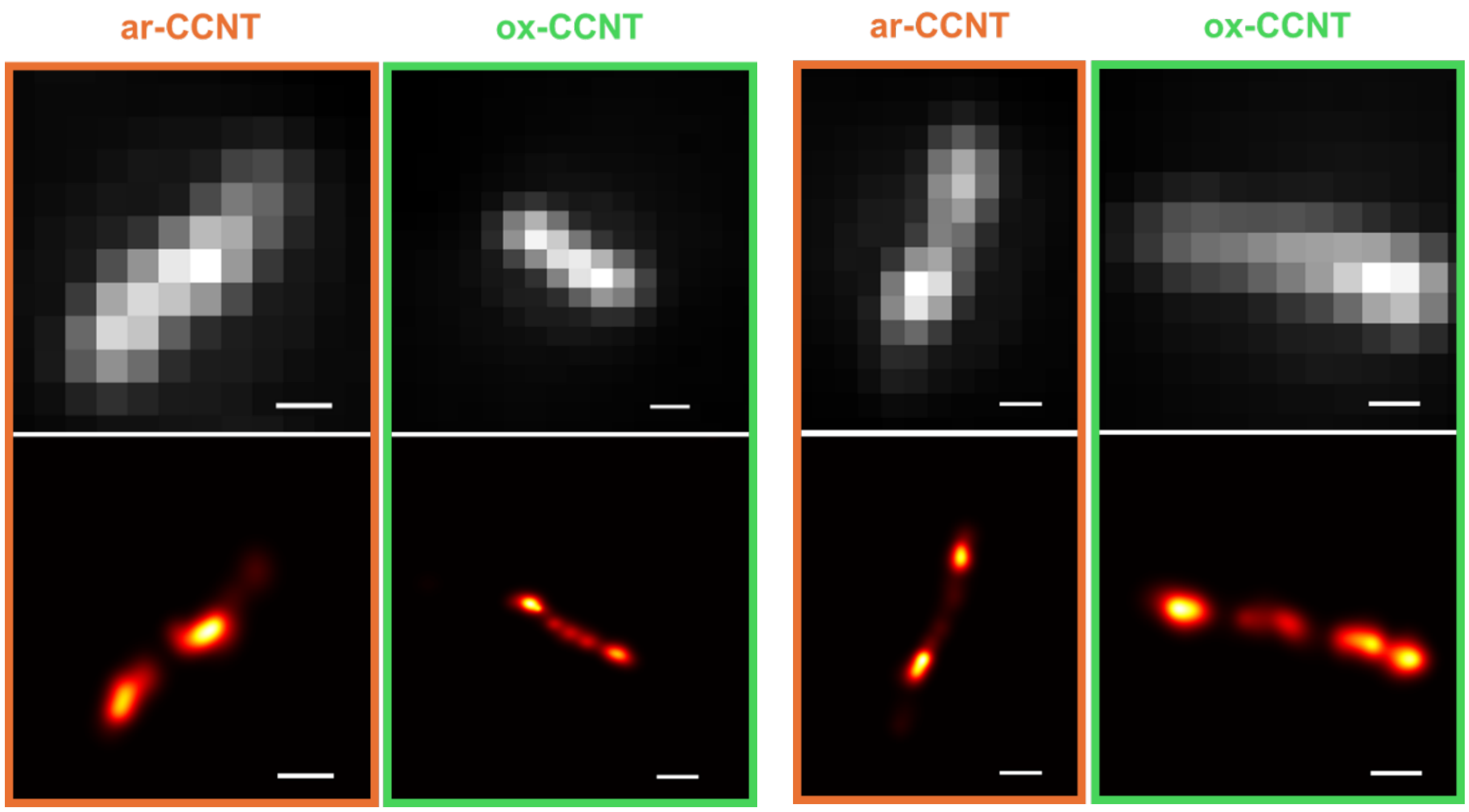

**Figure S12**. Super-resolved images for long ar-CCNT (orange) and ox-CCNT (green) showing non-fluorescent segments. Scale bars: 1 µm.

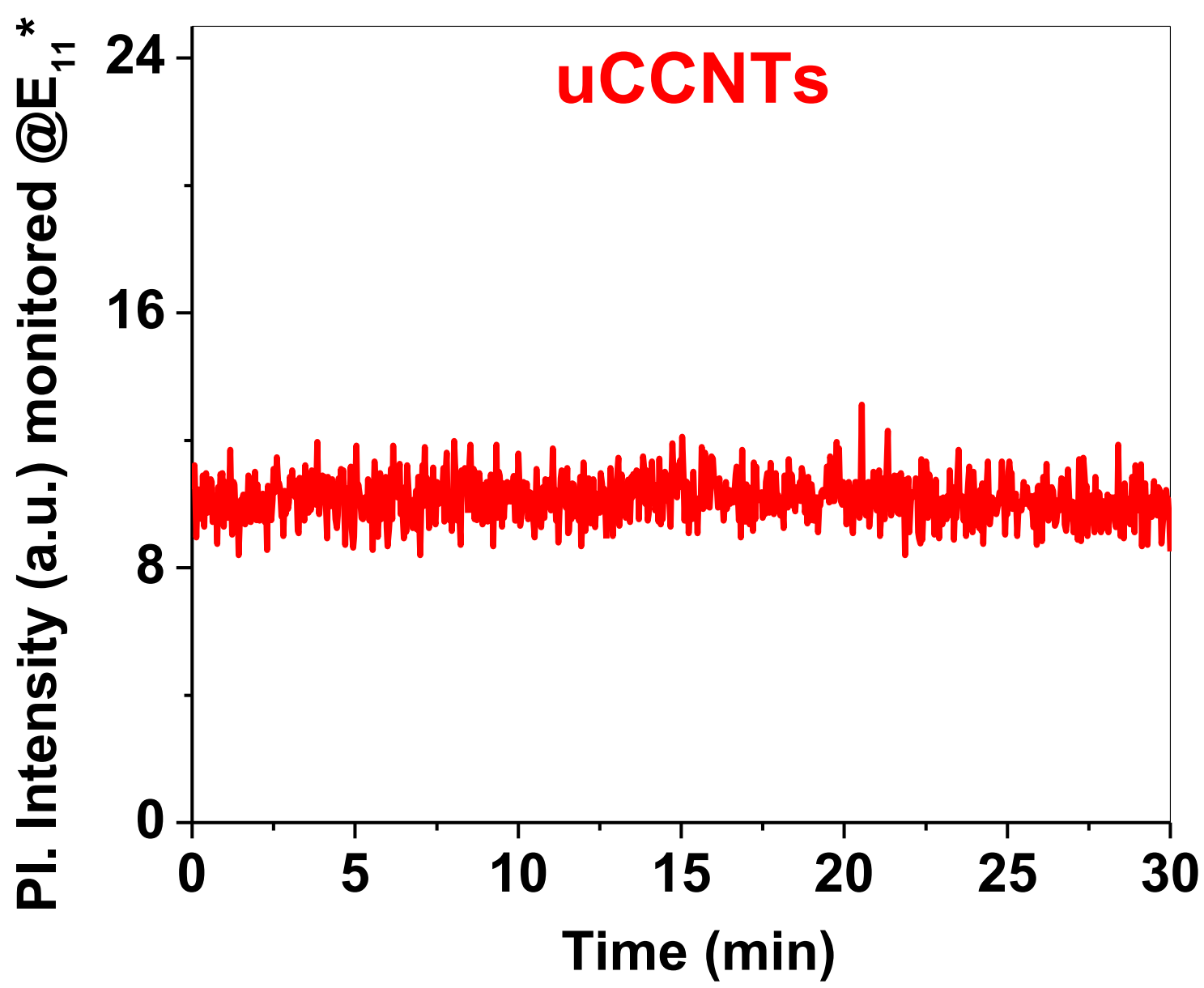

**Figure S13**. Photostability of ar-uCCNTs measured at the ensemble level by monitoring the $E_{11}$* PL intensity, showing that the uCCNTs are extremely photostable.

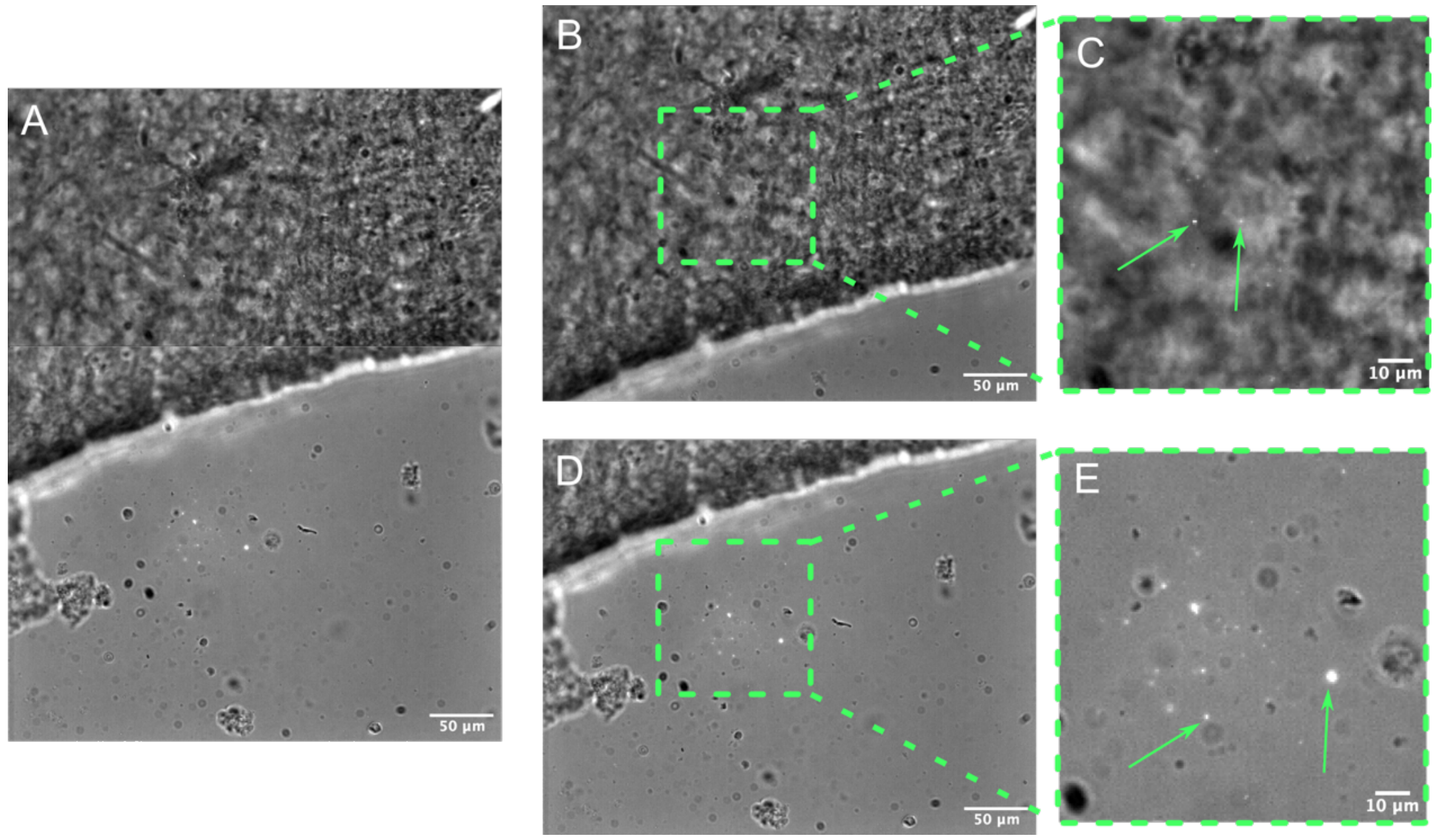

**Figure S14. Ex vivo NIR imaging of ox-uCCNTs through mouse brain slices.** A) Composite transmission-bright field overview of a 50-µm thick brain slice mounted on a glass coverslip bearing immobilized ox-uCCNTs imaged by NIR epifluorescence. B) details of the same field-of-view, showing bright, discrete emission spots from ox-uCCNTs located directly beneath the tissue (green box outlines region enlarged in C; 985 nm excitation). C) Magnified inset of (B), with green arrows marking individual ox-uCCNTs. D) NIR image of ox-uCCNTs positioned at the periphery of the slice (green box outlines region enlarged in E; scale bar: 50 µm), serving as a control to show the absence of brain autofluorescence. E) Close-up of the peripheral region in (D), with arrows highlighting ox-uCCNT emission spots.

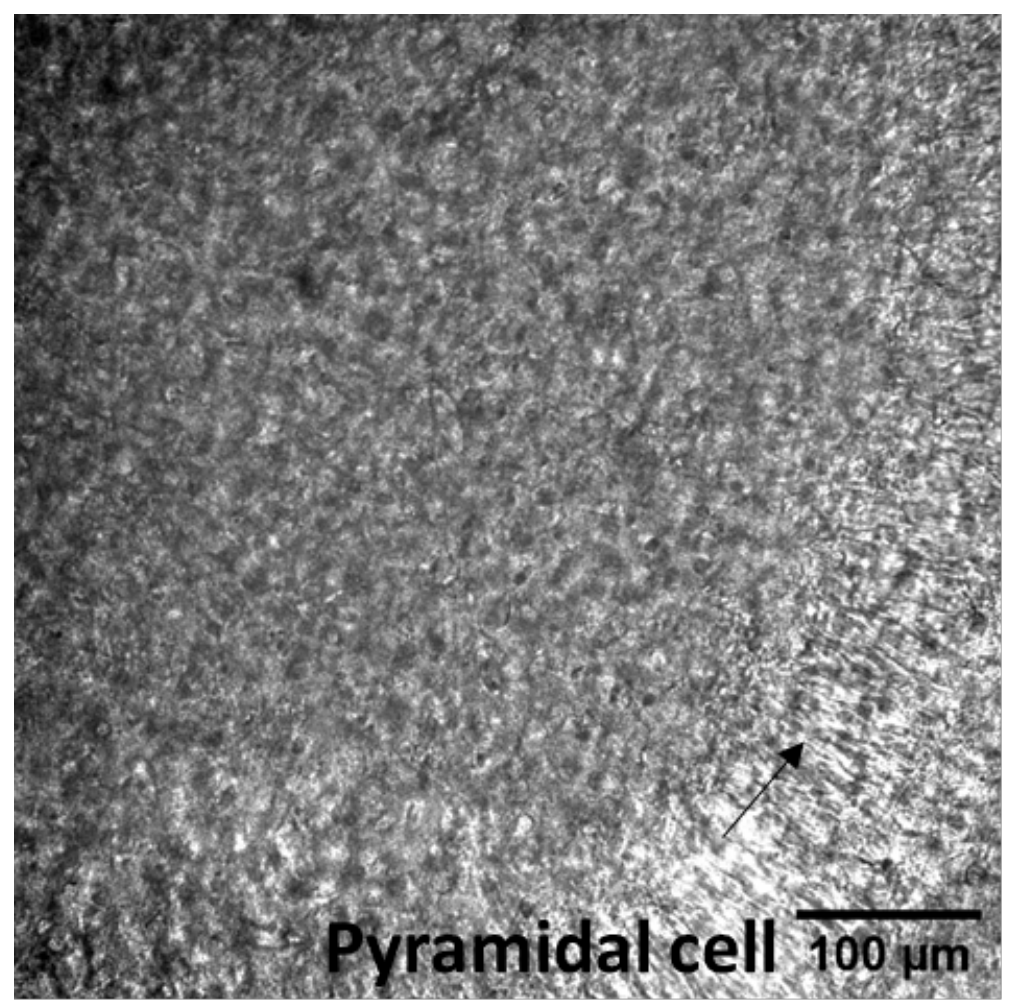

**Figure S15**. Tissue integrity of organotypic slices through oblique illumination (with a white light, Intensilight, Nikon).

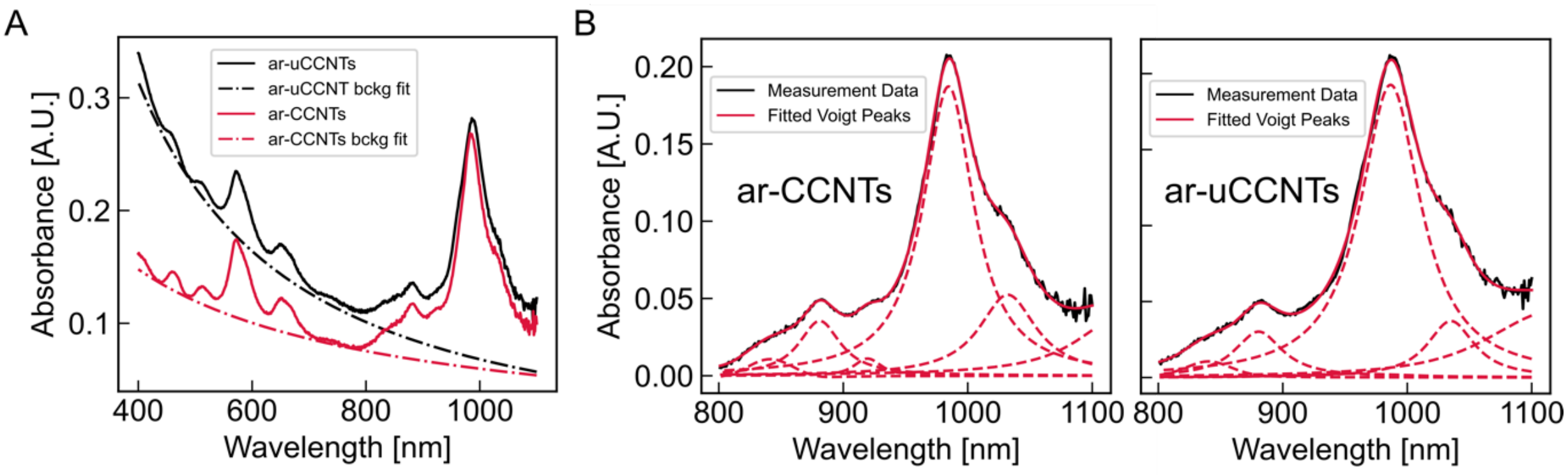

**Figure S16**. Determination of $E_{11}$ absorbance values for ar-CCNTs and ar-uCCNTs in PLPEG solutions. (A) Fitting of scattering background from absorbance spectra. (B) Measurement of $E_{11}$ absorbance peaks of (6,5) ar-(u)CCNTs dispersions (main peak in Figure B).

**Determination of PLQYs**

**Table S1. Photoluminescence quantum yields (PLQYs) obtained from ensemble measurement for four different samples**. The typical relative error of PLQY measurements is estimated to be 10-15%, mainly due to errors in the absorption measurement.

| **Samples** | **PLQYs (%)** | | |
|---|---|---|---|
| | **$E_{11}$** | **$E_{11}$*** | Total |
| ar-uCCNTs | 0.05 | **0.32** | 0.37 |
| ar-CCNTs | 0.03 | **0.26** | 0.29 |
| ox-uCCNTs | 0.09 | **0.70** | 0.79 |
| ox-CCNTs | 0.21 | **2.12** | 2.33 |

**Movie:**

**Supplementary Movie S1**. 3D single-particle tracking of uCCNTs in organotypic slice using DH-PSF.